\documentclass[preprintnumbers,article,amsmath,amssymb,floatfix,10pt,prd,onecolumn,superscriptaddress,nofootinbib]{revtex4-2}

\usepackage{titlesec}
\usepackage[toc]{appendix}
\usepackage{caption}
\titleformat*{\section}{\LARGE\bfseries}
\titleformat*{\subsection}{\Large\bfseries}
\titleformat*{\subsubsection}{\large\bfseries}
\titleformat*{\paragraph}{\large\bfseries}
\titleformat*{\subparagraph}{\large\bfseries}
\usepackage{bm}
\usepackage{amsfonts}
\usepackage{latexsym}
\usepackage{graphicx}
\usepackage{amsmath}
\usepackage{palatino}
\usepackage{mathpazo}
\usepackage{textcomp}
\usepackage{float}
\usepackage{booktabs}
\usepackage{dcolumn}
\usepackage{ragged2e}
\usepackage{hyperref}
\hypersetup{colorlinks,citecolor=blue}
\hypersetup{colorlinks=true,linkcolor=blue,filecolor=magenta,    urlcolor=blue}
\usepackage{amsthm}
\usepackage{xcolor}
\usepackage{orcidlink}
\usepackage{epsfig}
\usepackage{subcaption}
\usepackage{commath}
\usepackage{cancel, ulem}
\usepackage[utf8]{inputenc}
\usepackage{multirow}
\def\jnl@style{\it}
\def\aaref@jnl#1{{\jnl@style#1}}

\def\aaref@jnl#1{{\jnl@style#1}}

\def\aj{\aaref@jnl{AJ}}                   
\def\apj{\aaref@jnl{ApJ}}                 
\def\apjl{\aaref@jnl{ApJ}}                
\def\apjs{\aaref@jnl{ApJS}}               
\def\apss{\aaref@jnl{Ap\&SS}}             
\def\aap{\aaref@jnl{A\&A}}                
\def\aapr{\aaref@jnl{A\&A~Rev.}}          
\def\aaps{\aaref@jnl{A\&AS}}              
\def\mnras{\aaref@jnl{Mon.~Not.~Roy.~Astron.~Soc.}}             
\def\prd{\aaref@jnl{Phys.~Rev.~D}}        
\def\prc{\aaref@jnl{Phys.~Rev.~C}}  
\def\prl{\aaref@jnl{Phys.~Rev.~Lett.}}    
\def\qjras{\aaref@jnl{QJRAS}}             
\def\skytel{\aaref@jnl{S\&T}}             
\def\ssr{\aaref@jnl{Space~Sci.~Rev.}}     
\def\zap{\aaref@jnl{ZAp}}                 
\def\nat{\aaref@jnl{Nature}}              
\def\aplett{\aaref@jnl{Astrophys.~Lett.}} 
\def\apspr{\aaref@jnl{Astrophys.~Space~Phys.~Res.}} 
\def\physrep{\aaref@jnl{Phys.~Rep.}}      
\def\physscr{\aaref@jnl{Phys.~Scr}}       
\def\commat{\aaref@jnl{Comm.~Math.~Phys.}}              
\def\science{\aaref@jnl{Science}}               
\def\cqg{\aaref@jnl{Classical Quant.~Grav.}}            
\def\jpcs{\aaref@jnl{JPCS}}                                     
\def\ijmpd{\aaref@jnl{Int.~J.~Mod.~Phys.~D}}                    
\def\grg{\aaref@jnl{Gen.~Relat.~Gravit.}}               
\def\rpp{\aaref@jnl{Rep.~Prog.~Phys.}}          
\def\npa{\aaref@jnl{Nucl.~Phys.~A}}        
\def\lrr{\aaref@jnl{Living Rev.~Rel.}}                   
\def\jcap{\aaref@jnl{J.~Cosmology Astropart.~Phys.}}    
\def\rmp{\aaref@jnl{Rev.~Mod.~Phys.}}   
\def\epjc{\aaref@jnl{Eur.~Phys.~J.~C}} 
\def\plb{\aaref@jnl{~Phy.~Lett.~B}} 
\def\mpla{\aaref@jnl{Mod.~Phy.~Lett.~A}} 
\def\arxiv{\aaref@jnl{arxiv.org}}

\allowdisplaybreaks[1]

\begin{document}

\title{A generic dynamical system formulation for Bianchi-I cosmology with isotropic fluid in $f(Q)$ gravity}
\author{Ghulam Murtaza\orcidlink{0009-0002-6086-7346}}
\email{ghulammurtaza@1utar.my}
\affiliation{Department of Mathematical and Actuarial Sciences, Universiti Tunku Abdul Rahman, Jalan Sungai Long,
43000 Cheras, Malaysia}
\author{Saikat  Chakraborty\orcidlink{0000-0002-5472-304X}}
\email{saikat.c@chula.ac.th}
\affiliation{High Energy Physics Theory Group, Department of Physics,\\
Faculty of Science, Chulalongkorn University, Bangkok 10330, Thailand}
\affiliation{Center for Space Research, North-West University, Potchefstroom 2520, South Africa}
\author{Avik De\orcidlink{0000-0001-6475-3085}}
\email{avikde@um.edu.my}
\affiliation{Institute of Mathematical Sciences, Faculty of Science, Universiti Malaya, 50603 Kuala Lumpur, Malaysia}


\footnotetext{The research has been carried out under Universiti Tunku Abdul Rahman Research Fund project IPSR/RMC/UTARRF/2023-C1/A09 provided by Universiti Tunku Abdul Rahman. S.C. is supported by the Second Century Fund (C2F), Chulalongkorn University, Thailand.}

\begin{abstract}

In this article, we present a generic dynamical system formulation for Bianchi-I cosmology in the presence of an isotropic fluid within the coincident gauge connection branch and one of the non-coincident gauge connection branches of $f(Q)$ gravity theory. For both the connection branches under consideration, we start from the generic Bianchi-I cosmological field equations in $f(Q)$ and present a prescription of how one can construct an autonomous dynamical system in terms of the standard Hubble-normalized dimensionless dynamical variables once an $f(Q)$ theory is provided. Particular care has been taken to single out the physically viable regions in the phase space for each of the models under consideration. This results in the finding that, for both of the connection branches under consideration, the Kasner solution marginally violates the key physical viability condition of positive effective gravitational coupling ($f_Q>0$) for all the models considered, whereas a physically viable de-Sitter future attractor appears in all the models, except for the very special case of the monomial model within the coincident gauge connection. In the context of the early universe cosmology, we find that isotropization of a homogeneously perturbed inflating FLRW universe is a generic model-independent feature in the coincident gauge, whereas the isotropization of a homogeneously perturbed pre-bounce ekpyrotically contracting FLRW universe is, although not completely generic, but a likely scenario.
\end{abstract}

\maketitle
\tableofcontents
\section{Introduction}\label{sec1}
According to the cosmological principle, homogeneity and isotropy define our universe, and it is characterized by the Friedmann-Lema\^itre-Robertson-Walker (FLRW) geometry. Tensions in the physical parameters arise from the cosmological observations for the early and late time phases of the universe being analyzed, which motivated the study of alternatives to the standard General Relativistic $\Lambda$CDM model. Such alternative scenarios may include gravity theories beyond Einstein's General Relativity (GR) and/or challenge the cosmological principle \cite{a1}. The presence of anisotropies and inhomogeneities in the CMB \cite{a6} suggests that the cosmological principle was broken in the early stages of the universe, which has led to the investigation of anisotropic and inhomogeneous cosmological solutions \cite{a2,a3,a4}.      

Here we consider the simplest extension of the FLRW universe, the anisotropic yet homogeneous Bianchi type I spacetime model, given by the line element
\begin{equation}\label{metric}
ds^2=-dt^2+a_1^2(t)dx^2+a_2^2(t)dy^2+a_3^2(t)dz^2.
\end{equation}
where $a_1(t)$, $a_2(t)$ and $a_3(t)$ are the different scale factors corresponding to the $x$, $y$ and $z$ principle axes, respectively. This formulation allows different expansion factors of the universe in three orthogonal directions. This makes the Bianchi-I metric a particularly good option for analyzing different cosmological models that deviate from perfect isotropy, as has been studied in \cite{5,6,7,8,10,11,12,14,16,17,19,20}.

 With the increasing efficiency in observational cosmology, it is quite established that general relativity (GR) cannot explain the late-time accelerating expansion of the universe without assuming the so-called dark energy (DE) in its dynamics. However, to date, the searches for this dark component have not produced any fruitful results regarding its existence. Moreover, in the framework of GR some observational \cite{23,24,21,22} and theoretical \cite{25,26,27} issues including $\sigma_{8}$ tension \cite{28,29}, Hubble tension \cite{28,30,a1} and the coincident problem \cite{32,33} are still unresolved.  Now, one way to overcome these limitations of the GR is to change its gravitational component, hence called modified theories of gravity. One of the simplest modifications of GR is $f(\mathring{R})$ theory in which the Ricci scalar $\mathring{R}$ of the Levi-Civita connection $\mathring{\Gamma}$ is replaced by an arbitrary, viable functional form $f(\mathring{R})$ in the Einstein-Hilbert action term \cite{34,35,36}. The latest trend in the modified gravity theories is to attribute gravity to torsion or non-metricity characteristic of the spacetime instead of curvature. This opens up a whole new avenue of looking towards gravitational interaction. The metric teleparallel and symmetric teleparallel theories in the respective directions and their extensions $f(T)$ \cite{surveyfT} and $f(Q)$ \cite{surveyfQ} theories are quite successful in explaining the latest observational aspects. 
 
 $f(Q)$ gravity theory can be constructed by replacing the torsion-free and metric-compatible Levi-Civita connection with an affine connection with vanishing curvature and vanishing torsion and letting its non-metricity control the gravitation, where $Q$ is the non-metricity scalar. Several cosmological studies have been done on $f(Q)$ theory in the past few years \cite{50,51,52,53,49,55,56,59,60,61,62,63,3}. However, most of these works are conducted in homogeneous and isotropic backgrounds, and only limited attention was drawn to the anisotropic universe in this theory. The theory was first implemented in the Bianchi type I anisotropic universe in \cite{15,64}, and subsequently in some other notable studies; see \cite{9, Paliathanasis2024, Leon2024} and the references therein. This interest in anisotropy follows previous works in the context of $f(T)$ gravity theory under metric teleparallelism, as presented in \cite{sharifftBI, AslamftBI, RodriguesftBI, TretyakovftBI, ColeyftBI}. We should mention here that some instabilities within $f(Q)$ have been revealed in \cite{R6}, where cosmological perturbations exhibit pathological behavior. 
 However, in \cite{R7}, it argued that 
 by reformulating the theory within the higher-order scalar-tensor framework and replacing scalar St\"uckelberg fields with vector fields, it was demonstrated that the second-class constraints in the Arnowitt-Deser-Misner formalism can suppress this ghost mode.

The Einstein field equations are partial differential equations (PDEs). However, by imposing spatially homogeneous space-time conditions, they reduce to ordinary differential equations (ODEs). In this setting, one can use the dynamical system approach, which proves to be a powerful tool for analyzing the qualitative behavior of the solutions. This has been particularly useful in cosmology \cite{66,67}, as it enhances our understanding of the universe's evolution, various cosmological phenomena, and modified theories of gravity \cite{68}. Notable advancements have been achieved using dynamical system analysis (DSA), such as a comprehensive study of $f(\mathring{R})$ gravity \cite{69} and stability analyses of various cosmological models within this framework \cite{70}. Additionally, finite-time singularities in $f(\mathring{R})$ gravity and interacting multifluid cosmology have been explored \cite{71}. A note on the formulations of the dynamical system in $f(\mathring{R})$ theory can be found in \cite{81}. DSA of two accelerating models of $f(T)$ theory, which generalizes the teleparallel equivalent of general relativity (TEGR) has been given in \cite{77}. DSA of torsion $T$ and the scalar field $\phi$ in two models of teleparallel theory has been conducted \cite{78}. DSA has also been studied to constrain $f(T)$ gravity \cite{79}.

There have been quite a few works on DSA of spatially homogeneous and isotropic FLRW cosmology in $f(Q)$ gravity. A detailed analysis of the power exponential model $f(Q) = Q e ^{\lambda \frac{Q_{0}}{Q}}$ using the DSA in $f(Q)$ theory has been carried out, considering two fluid components, radiation, and matter in \cite{72}. For the model parameter $\lambda \not =0$, the existence of a radiation-dominated early time epoch, a saddle point dominated by matter, and an accelerating de Sitter attractor has been found. DSA of the background and perturbations in $f(Q)$ gravity was performed in \cite{73}, for both the exponential and the power-law models of $f(Q)$. In both cases, the analysis revealed a matter-dominated saddle point with the correct growth rate of matter perturbations. This phase is followed by a smooth transition to a stable, dark energy-dominated accelerated universe, where matter perturbations remain constant. DSA of two models of $f(Q)$, the power-law model $f(Q) = Q+mQ^{n}$ and logarithmic model $f(Q) = \alpha + \beta \log Q$ has been carried out in \cite{74}. Both models lead to an accelerating and stable universe characterized by constant matter perturbations. Additionally, the power-law model results in a matter-dominated saddle point with the correct matter perturbation growth rate. In contrast, the logarithmic model leads to a saddle point dominated by the geometric component of the model, with perturbations in the matter sector. DSA of scalar field cosmology in coincident $f(Q)$ gravity has been done in \cite{75}. Several other recent and significant studies in $f(Q)$ theory using DSA approach are presented in \cite{Channuie2024,solanki2024,Mahataa2024,narawade2023,alam2023,4}.

It is known that spatially flat FLRW cosmology admits three symmetric teleparallel curvatureless torsionless connection branches \cite{Hohmann:2021ast}, which can be denoted as $\Gamma_1,\Gamma_2,\Gamma_3$. Most of the works on FLRW DSA in $f(Q)$ that are mentioned in the last paragraph deal with the first connection branch, which, in the context of spatially flat FLRW cosmology, is also called the ``coincident gauge'' \footnote{There is a slight misnomer here. For any particular connection branch $\Gamma_i$, one can find a coordinate system in which the connection can be trivialized \cite{Hohmann:2021ast}. This particular coordinate system is the coincident gauge of the connection branch in question. It happens that the coincident gauge for the connection branch $\Gamma_1$ is the usual Cartesian coordinate system.}. A generic dynamical system for the connection branch $\Gamma_1$ in $f(Q)$ gravity was provided by Boehmer et.al. in \cite{Boehmer:2022wln}, who also established for the first time that the FLRW cosmological phase space for the connection branch $\Gamma_1$ in the presence of a single fluid will be 1-dimensional. Not much work on the $f(Q)$ DSA with the connection branches $\Gamma_2$ and $\Gamma_3$ are found in the literature. Paliathanasis \cite{Paliatha2023} has worked out the DSA for the simple monomial $f(Q)\propto Q^n$ form for $\Gamma_2,\Gamma_3$. Shabani et.al. \cite{80} appear to have presented for the first time a generic dynamical system formulation for $\Gamma_2$. In \cite{Shabani2024}, a generalized class of the $\Gamma_3$ branch was utilized to analyze the role of the spatial curvature in FLRW spacetime. 
In \cite{Palia2025}, analysis of various cosmological observational datasets was conducted within the framework of non-coincident $f(Q)$ gravity, focusing on a power-law model. The results offer an improved fit to the data compared to the standard $\Lambda$CDM model. Similar, the analysis presented in \cite{55,56} demonstrated that the $f(Q)$ gravity, formulated within the coincident gauge provides a better fit to certain observational datasets than the $\Lambda$CDM model.

In comparison to FLRW, fewer studies have explored the DSA of the anisotropic spacetime models in the context of $f(Q)$ gravity. Recently, DSA of the LRS Bianchi-I universe in $f(Q)$ theory was carried out in the papers \cite{2, Rathore:2024ema}. However, both of these works consider a special case where two spatial directions evolve identically, thus simplifying their analysis. In \cite{Leon2024}, the authors investigated the Bianchi-I spacetime in $f(Q)$ theory starting from a minisuperspace description and then constructing a dynamical system. The authors put particular emphasis on delving deep into the dynamics of Kasner and Kasner-like solutions, analyzing the effects of different connections, especially in regimes where the nonmetricity scalar $Q$ vanishes. Another work in which DSA of Bianchi-I cosmology has been explored is \cite{Esposito:2022omp}, in which the authors also consider an example with anisotropic pressure. Applications of non-coincident gauges in other Bianchi-type spacetimes within the framework of $f(Q)$ gravity can be found in the literature. For instance, in \cite{Roumeliotis2023}, the authors derived exact self-similar solutions for specific functional forms of $f(Q)$ in both vacuum and perfect fluid cases by employing two distinct flat, symmetric connections (i.e., non-coincident gauges) for the Kantowski–Sachs and Bianchi III LRS metrics. Furthermore, \cite{Millano2024} demonstrated that these solutions arise only when the connection is non-trivial, i.e., beyond the coincident (zero-connection) gauge, emphasizing that the choice of gauge in symmetric teleparallel gravity has a direct impact on the physical cosmological dynamics.

What was lacking in all the above-mentioned studies of $f(Q)$ DSA in Bianchi-I is a rigorous generic dynamical system formulation in the presence of matter. For a generic dynamical system formulation, one must present a unified prescription such that, given any $f(Q)$, one is able to construct an autonomous dynamical system. 
At this point, we acknowledge that the recent work \cite{Leon2024} has attempted to provide a generic dynamical system formulation for Bianchi-I cosmology across all the connection branches in the absence of matter, adopting a minusuperspace approach and resorting to an effective scalar field description. In contrast, the generic dynamical system formulation presented in this work does not require one to go to the effective scalar field description. In that sense, our work can be considered an extension to the generic dynamical system formulation for $f(Q)$ as presented by Boehmer et.al. \cite{Boehmer:2022wln}, and parallel to the work \cite{Chakraborty:2018bxh} where a generic dynamical system formulation was provided for Bianchi-I cosmology in $f(R)$ gravity. Apart from the obvious difference presented by the inclusion of matter into the picture, two other key differences of our work as compared to \cite{Leon2024} are the reduction of the phase space dimensionality, and a clear emphasis to single out the physically allowed region of the phase space. We believe our work will serve as a starting point for more generic analyses of the anisotropic universe within the $f(Q)$ framework.

This paper is organized as follows. In section \ref{sec2}, we discuss the basic formalism of $f(Q)$ gravity. In section \ref{sec3}, we derive the Bianchi-I cosmology in $f(Q)$ theory. In section \ref{sec4}, we derive the general dynamical system for  Bianchi-I cosmology in $f(Q)$ theory for $\Gamma_1$. In section \ref{sec5}, we apply our generic dynamical system formulation to late-time cosmology for $\Gamma_1$. In section \ref{dsaII}, we derive the general dynamical system for  Bianchi-I cosmology in $f(Q)$ theory for $\Gamma_2$. In section \ref{lateG2}, we apply our generic dynamical system formulation to late-time cosmology for $\Gamma_2$.  In section \ref{sec6}, we discuss the application of early universe cosmology.  In section \ref{comptfTG}, we provide a brief comparison with $f(T)$ gravity. We summarise our findings in the conclusion section \ref{sec7}.

\section{Basic Formalism of $f(Q)$-Gravity}\label{sec2}
In $f(Q)$-gravity theory, the spacetime is constructed by using the symmetric teleparallelism and non-metricity condition, that is, $R^\rho{}_{\sigma\mu\nu} = 0$ and $Q_{\lambda\mu\nu} := \nabla_\lambda g_{\mu\nu} \neq 0$. The associated connection coefficient is given by 
\begin{equation} \label{connc}
\Gamma^\lambda{}_{\mu\nu} = \mathring{\Gamma}^\lambda{}_{\mu\nu} +L^\lambda{}_{\mu\nu}
\end{equation}
where $\mathring{\Gamma}^\lambda{}_{\mu\nu}$ is the Levi-Civita connection and $L^\lambda{}_{\mu\nu}$ is the disformation tensor. This implies that 
\begin{equation*}
L^\lambda{}_{\mu\nu} = \frac{1}{2} (Q^\lambda{}_{\mu\nu} - Q_\mu{}^\lambda{}_\nu - Q_\nu{}^\lambda{}_\mu) \,.
\end{equation*}
In addition, we define the superpotential tensor
\begin{equation} \label{P}
P^\lambda{}_{\mu\nu} := \frac{1}{4} \left( -2 L^\lambda{}_{\mu\nu} + Q^\lambda g_{\mu\nu} - \tilde{Q}^\lambda g_{\mu\nu} -\frac{1}{2} \delta^\lambda_\mu Q_{\nu} - \frac{1}{2} \delta^\lambda_\nu Q_{\mu} \right) ,\,
\end{equation}
and using it, the non-metricity scalar 
\begin{equation} \label{Q}
Q = Q_{\lambda\mu\nu}P^{\lambda\mu\nu} = -\frac{1}{2}Q_{\lambda\mu\nu}L^{\lambda\mu\nu} + \frac{1}{4}Q_\lambda Q^\lambda - \frac{1}{2}Q_\lambda \tilde{Q}^\lambda \,. 
\end{equation}
The action of $f(Q)$-gravity is given by
\begin{equation*}
S = \int \left[\frac{1}{2\kappa}f(Q) + \mathcal{L}_M \right] \sqrt{-g}\,d^4 x
\end{equation*}
where $g$ is the determinant of the metric tensor and $\mathcal{L}$ is the matter Lagrangian. By varying the action with respect to the metric, we obtain
\begin{equation} \label{FE1}
\frac{2}{\sqrt{-g}} \nabla_\lambda (\sqrt{-g}f_QP^\lambda{}_{\mu\nu}) -\frac{1}{2}f g_{\mu\nu} + f_Q(P_{\nu\rho\sigma} Q_\mu{}^{\rho\sigma} -2P_{\rho\sigma\mu}Q^{\rho\sigma}{}_\nu) = \kappa T_{\mu\nu}.
\end{equation}
 Nevertheless, this equation is not in a tensor form, and it is only valid in the coincident gauge coordinate \cite{Jimenez/2018}. 

On the other hand, by using (\ref{connc}), we can have the following relations between the curvature tensors corresponding to $\Gamma$ and $\mathring{\Gamma}$:
\begin{equation}
R^\rho{}_{\sigma\mu\nu} = \mathring{R}^\rho{}_{\sigma\mu\nu} + \mathring{\nabla}_\mu L^\rho{}_{\nu\sigma} - \mathring{\nabla}_\nu L^\rho{}_{\mu\sigma} + L^\rho{}_{\mu\lambda}L^\lambda{}_{\nu\sigma} - L^\rho{}_{\nu\lambda} L^\lambda{}_{\mu\sigma}
\end{equation}
and so
\begin{align*}
R_{\sigma\nu} &= \mathring{R}_{\sigma\nu} + \frac{1}{2} \mathring{\nabla}_\nu  Q_\sigma + \mathring{\nabla}_\rho L^\rho{}_{\nu\sigma} -\frac{1}{2} Q_\lambda L^\lambda{}_{\nu\sigma} - L^\rho{}_{\nu\lambda}L^\lambda{}_{\rho\sigma} \nonumber \\
R &= \mathring{R} + \mathring{\nabla}_\lambda Q^\lambda - \mathring{\nabla}_\lambda \tilde{Q}^\lambda -\frac{1}{4}Q_\lambda Q^\lambda +\frac{1}{2} Q_\lambda \tilde{Q}^\lambda - L_{\rho\nu\lambda}L^{\lambda\rho\nu} \,.
\end{align*}
Therefore, by using the symmetric teleparallelism condition, we can rewrite the field equations in (\ref{FE1}) as
\begin{equation} \label{FE2}
f_Q \mathring{G}_{\mu\nu} + \frac{1}{2}g_{\mu\nu}(Qf_Q - f) + 2f_{QQ} \mathring{\nabla}_\lambda Q P^\lambda{}_{\mu\nu} = \kappa T_{\mu\nu} .
\end{equation}
where $$\mathring{G}_{\mu\nu} = \mathring{R}_{\mu\nu} - \frac{1}{2} g_{\mu\nu} \mathring{R}$$ and $T_{\mu\nu}$ is the energy-momentum tensor.

\section{Bianchi-I cosmology in $f(Q)$ gravity}\label{sec3}

For the current scenario, in the anisotropic Bianchi-I metric in Cartesian coordinates given by (\ref{metric}), the directional Hubble parameters are given by $H_i=\frac{\dot{a_i}}{a_i}$. We denote the arithmetic mean of these directional Hubble parameters by 
\begin{align}
    H(t)=\frac 13\left[H_1+H_2+H_3\right].
\end{align}
We also denote by $a(t)$, the average scale factor as the geometric mean
\begin{align}
    a(t)=\left[a_1(t)a_2(t)a_3(t)\right]^{\frac13}.
\end{align}
if we use the parametrization 
\begin{align}
    a_i(t)=a(t)e^{\beta_i(t)},
\end{align}
then we obtain \begin{align}
    H_i=H+\dot{\beta}_i, \quad \beta_1+\beta_2+\beta_3=0.
\end{align}
We denote the anisotropy parameter by
\begin{align}{\label{sigmaeq}}
\sigma^2=-2\sum \dot{\beta_i}\dot{\beta_j}=\sum \dot{\beta_j^2}.
\end{align}
Note that $\sigma=0$ implies $\sum \dot{\beta_j^2}=0$, which implies that the universe is an isotropic FLRW. 

To study non-trivial isotropization in the evolution process of the universe, once the anisotropic type spacetime metric is considered, the EoS parameter of the gravitational fluid should, in principle, also be generalized to exhibit an anisotropic character to give a more sensible model. With the isotropization of the universe, the fluid also isotropizes to display a vanishing skewness parameter and isotropic pressure. The energy-momentum tensor for the anisotropic fluid is defined as
\begin{equation} \label{T}
T^{\nu}_{\mu} = \text{diag}(-\rho,P_1,P_2,P_3)=\text{diag}(-\rho, \omega_1 \rho,\omega_2 \rho,\omega_3 \rho). \,
\end{equation}
where $\rho$ denotes the energy density of the fluid, $P_1,\,P_2,\,P_3$ are the pressures along the three orthogonal directions which assume respective directional Equation of state (EoS) parameters $\omega_1,\,\omega_2,\,\omega_3$. One can define an average equation of state parameter $\omega$ and the deviations $\mu_i$ from the average equation of state as follows,
\begin{equation}
    \omega = \frac{1}{3}(\omega_1 + \omega_2 + \omega_3), \quad \mu_i = \omega_i - \omega, \quad \omega_i = \omega + \mu_i,
\end{equation}
for $i=1,2,3$.

In the subsequent part of the paper, we confine ourselves to the case of the isotropic fluid $P_1=P_2=P_3=P=\omega\rho$. One may argue, and justifiably so, about the justification of having an anisotropy in the geometry sector when there is no anisotropy in the fluid sector. Although such a situation is frequently explored in the literature, strictly speaking, this can be true only when the anisotropy is perturbatively small. This issue is explained in the appendix \ref{app}.

There are three isometries, the vector fields $\partial_x$, $\partial_y$ and $\partial_z$. Therefore, in terms of a dynamic degree of freedom $\gamma(t)$, the three symmetric and flat connection classes describe the Bianchi-I geometry given in Cartesian coordinates as \cite{Leon2024}. 
\begin{itemize}
    \item $\Gamma_1: \Gamma^t_{\,\,\, tt}=\gamma(t)$
    \item $\Gamma_2: \Gamma^t_{\,\, tt}=\gamma(t)+\frac{\dot \gamma(t)}{\gamma(t)},\quad \Gamma^t_{\,\, ti}=\gamma(t)$
    \item $\Gamma_3: \Gamma^t_{\,\, tt}=-\frac{\dot \gamma(t)}{\gamma(t)},\quad \Gamma^t_{\,\, ti}=-\gamma(t)$ 
\end{itemize}

\section{General formulation of the Dynamical system for $\Gamma_I$}\label{sec4}
In this section, we consider the $f(Q)$ gravity formulation in anisotropic Bianchi-I spacetime considering the first connection class $\Gamma_1$. We provide a stepwise prescription on how to construct a generic autonomous system, given any $f(Q)$ gravity model. Starting from an arbitrary differentiable function $f(Q)$, we first isolate a single transparent invertibility condition that guarantees system closure and thereafter provide an algorithm that turns field equations into a fully autonomous system. The domain of validity and the point at which the procedure fails are explicitly stated. 
We can derive the non-metricity scalar $Q$ as
\begin{equation}\label{Q}
-Q=2\sum H_iH_j=6H^2-\sigma^2.
\end{equation}



The modified Friedmann and Raychaudhuri equation for this connection class can be expressed as

\begin{align}
3H^2-\sigma^2 =\frac\kappa{f_Q}\left[ \rho-\frac{3H^2f_Q}{\kappa}-\frac f{2\kappa}\right]\,,
\end{align}
\begin{align}
-(2\dot H+3H^2)-\frac{\sigma^2}{2} = \frac{\kappa}{f_Q}\left[ \omega \rho+\frac f{2\kappa}+\frac{2H\dot{f_Q}}{\kappa}-\frac{Qf_Q}{2\kappa}  \right].
\end{align}

The two above equations can be further simplified, taking into account the expression of $Q$ from \eqref{Q}. Ultimately, the Bianchi-I cosmological field equations in the presence of an isotropic fluid in the context of $f(Q)$ gravity with the connection branch $\Gamma_1$ are
\begin{align}\label{1eq:m}
3H^2 - \frac{\sigma^2}{2} = \frac{\kappa}{2f_Q}\left[\rho - \frac{f}{2\kappa}\right], 
\end{align}
\begin{align}\label{2eq:m}   
-(\dot H+3H^2) = \frac{\kappa}{2f_Q}\left[\omega \rho + \frac{f}{2\kappa}+\frac{2H\dot{f_Q}}{\kappa}\right]. 
\end{align}
The anisotropy evolution equation is given by (see Appendix \ref{dotsigma} for details)
\begin{align}
   \dot{\sigma} = -\sigma\left(\frac{\dot{f}_{Q}}{f_{Q}}+3H\right) .
\end{align}
which can be integrated as
\begin{align}\label{3eq:m}
    \sigma \sim \frac{1}{a^3f_Q} .
\end{align}
In the absence of hypermomentum, i.e. absence of matter-connection coupling, the standard continuity relation holds
\begin{align}\label{conteq}
    \dot \rho +3H(p+\rho)=0.
\end{align}

Notice that for GR ($f_Q\to1$) it is easier to find how the anisotropy evolves; It decreases (increases) as $\sim\frac{1}{a^3}$ in an expanding (contracting) universe. It is not so straightforward to find the behaviour of anisotropy in the $f(Q)$ theory because of the existence of the $f_Q$ term in the denominator in \eqref{3eq:m}, since the expression for $Q$ itself involves $\sigma$. In general, \eqref{3eq:m} becomes a nonlinear equation for $\sigma$ and it is not possible to solve for it except for certain simple cases\footnote{A similar situation arises in $f(R)$ gravity \cite{Bhattacharya:2017cbn}.}. This is what motivates us to take up the dynamical system analysis in the later sections. In particular, this also gives the hope that some of the $f(Q)$ forms by themselves may tackle the anisotropy divergence problem in a pre-bounce ekpyrotic contraction phase, without needing to incorporate a superstiff fluid or a fast rolling scalar field (see a similar work in the context of $f(R)$ in \cite{Arora:2022dti}).

From the equations \eqref{1eq:m} and \eqref{2eq:m}, it can be seen that the effective gravitational coupling is $\frac{\kappa}{f_Q}$. To avoid an anti-gravity situation, one must demand $f_Q>0$, which is one of the physical viability conditions to be satisfied by an $f(Q)$ theory \cite{Guzman:2024cwa}.

We start by defining the dimensionless variables
\begin{align}
   x_{1} = \frac{\kappa \rho}{6f_{Q}H^{2}}, \qquad x_{2} = -\frac{f}{12f_{Q}H^{2}}, \qquad x_{3} = \frac{\sigma^{2}}{6H^{2}}.
   \label{variables}
\end{align}
constrained by the relation (\ref{1eq:m}) 
\begin{align} \label{constraint}
    1=x_{1}+x_{2}+x_{3}.
\end{align}
Now to construct the dynamical system equations, we take the derivative of the dimensionless variables $x_{1}$,~$x_{2}$ and $x_{3}$ with respect to cosmic time $t$ and use the Hubble normalized dimensionless time variable $N=\ln a$ to get the following,
\begin{align}
    x'_{1} = -3x_{1}(1+\omega)+6x_{1}\left[\frac{(x_{2}-1-x_{1}\omega)(1+\Gamma-\Gamma x_{3}-2x_{3})}{(\Gamma +2)(x_{3}-1)}\right]-\frac{6x_{1}x_{3}}{(\Gamma +2)(x_{3}-1)},
\end{align} 
\begin{align}
    x'_{2} = 3\Gamma \left(\frac{x_{3}+x_{2}-1-x_{1}\omega}{\Gamma +2}\right)+6x_{2}\left[\frac{(x_{2}-1-x_{1}\omega)(1+\Gamma-\Gamma x_{3}-2x_{3})}{(\Gamma +2)(x_{3}-1)}\right]-\frac{6x_{2}x_{3}}{(\Gamma +2)(x_{3}-1)},
\end{align}  
\begin{align}
  x'_{3} = 6x_{3}(x_{1}\omega -x_{2}).
\end{align}
In the above prime represents derivative with respect to $N$ and we have defined the auxiliary quantity $\Gamma = \Gamma(Q)$ as,
\begin{align}\label{Gamma}
\Gamma = \frac{f_{Q}}{Q f_{QQ}}.
\end{align}
Using the (\ref{2eq:m}), we have,
\begin{align}\label{cosmology}
\frac{\dot{H}}{H^{2}} = 3\left( \frac{(2x_{3}+x_{3}\Gamma -\Gamma)(x_{2}-1-x_{1}\omega )+2x_{3}}{(\Gamma+2)(-1+x_{3})}\right).
\end{align}
Using the constraint to eliminate the $x_{2}$, dynamical system equations reduce to the following way,
\begin{align}\label{gds1}
x'_{1} = -3x_{1}(1+\omega)+6x_{1}(x_{1}+x_{3}+x_{1}\omega)+6x_{1}\frac{x_{1}+x_{1}\omega}{(\Gamma +2)(x_{3}-1)},
\end{align}
\begin{align}\label{gds2}   
x'_{3} = 6x_{3}x_{1}(1+\omega)+6x^{2}_{3}-6x_{3},
\end{align}

while \eqref{cosmology} reduces to
\begin{align}\label{cosmology_new}
\frac{\dot{H}}{H^{2}} =  3\left( \frac{(2x_{3}+x_{3}\Gamma -\Gamma)(-x_{3}-x_{1}-x_{1}\omega )+2x_{3}}{(\Gamma+2)(-1+x_{3})}\right).
\end{align}

The above expression helps us determine the cosmic evolution for a particular fixed point.

To write an autonomous dynamical system for a particular given $f(Q)$ model, we need to express $Q$, and hence $\Gamma(Q)$ in terms of dynamical variables $x_{1}$ and $x_{3}$, for which we follow the following mechanism. From (\ref{Q}) and (\ref{variables}), we can express $Q$ as,
\begin{align}\label{Qvar}
    \frac{Q f_Q}{f} = \frac{(1-x_3)}{2x_2}.
\end{align}
Further using the constraint equation (\ref{constraint}), one can write 
\begin{align}\label{Qvar1}
    \frac{Q f_Q}{f} =  \frac{(1-x_3)}{2(1-x_1-x_3)}.
\end{align}
Provided the above equation is invertible to get $Q=Q(x_1,x_3)$, one can substitute that into the expression of $\Gamma(Q)$ to obtain an autonomous dynamical system for the connection branch $\Gamma_1$ for Bianchi-I cosmology with an isotropic fluid in $f(Q)$ gravity. Notice that the phase space is 2-dimensional, which is in line with the result of Boehmer et.al. in \cite{Boehmer:2022wln}. The result of \cite{Boehmer:2022wln} is that in the presence of a single perfect fluid, the FLRW phase space for $\Gamma_1$ in $f(Q)$ is 1-dimensional. Here we just add one extra dynamical degree of freedom in the geometry sector, which is the anisotropy scalar $\sigma$, increasing the phase space dimensionality by one. 

In the following section, we take four $f(Q)$ gravity models and conduct the dynamical system analysis of Bianchi-I cosmology in $f(Q)$. All of these models have shown up in the $f(Q)$ literature in the past, and we present the relevant references corresponding to each of the models. For this analysis, we find the critical points of the autonomous system for each model and perform their linear stability analysis. Existence conditions for each critical point, depending on the model parameters, are also discussed.

\section{Application in the context of late-time cosmology for $\Gamma_I$}\label{sec5}

In this section, we demonstrate the applicability of our generic dynamical system formulation by applying it explicitly to four $f(Q)$ models that are found in the literature. In this section, we perform the analysis in the presence of a matter component, and specialise to the case of dust ($\omega=0$) while plotting the phase portraits. Therefore, at the back of our mind, we have a late time cosmology. Given the tight constraint on the amount of spatial anisotropy in the present-day observable universe, we expect an asymptotic isotropization in the future. 

\subsection{Model-I: $f(Q) = \alpha(-Q)^n$ ($n\neq1$)}\label{model1}

As a first example, we consider the simplest power law model with constant $\alpha$ and $n$. Now solving $f_Q$ and $f_{QQ}$ for this specific model, one can obtain
\begin{align}
    f_Q=-n\alpha(-Q)^{n-1}, ~~ f_{QQ}=n(n-1)\alpha(-Q)^{n-2}.
    \label{fq5}
\end{align}
Explicit substitution of this specific model $f(Q)$ into (\ref{Qvar1}) gives us an additional constraint,
\begin{align}\label{anotherconstr}
    x_3=1-\frac{2nx_1}{2n-1},
\end{align}
Substituting (\ref{fq5}) into (\ref{Gamma}) we obtain,
\begin{align}
    \Gamma = \frac{1}{n-1}.
\end{align}
Substituting the above expression of $\Gamma$ and using the constraint (\ref{anotherconstr}) into (\ref{gds1}), we get the one dimensional dynamical system,
\begin{align}
   x'_{1} = -3x_{1}(1+\omega)+6x_{1}\left(x_{1}+x_{1}\omega+1-\frac{2nx_1}{2n-1}\right)-\frac{3x_1(1+\omega)(n-1)}{n}.
   \label{4MDSE1}
\end{align}
For this particular model \eqref{cosmology_new} gives
\begin{equation}\label{q_conn1}
    \frac{\dot H}{H^2} = \frac{3 \left(1 + \omega - 2 n^2 (-1 + x_1) \omega + 
   n \left(-2 + x_1 + (-3 + x_1) \omega \right)\right)}{n ( -1 + 2n)}.
\end{equation}
The fixed points and their stabilities are listed in the Tables \ref{tablenew1} and \ref{tablenew2}.
\begin{table}[H]
\centering
\begin{tabular}{|p{2cm}|p{3.3cm}|p{2cm}|p{5.3cm}|}
\hline
\textbf{Fixed Points} & \textbf{Coordinates $x_1$} & \textbf{Existence} & \textbf{Eigenvalues} \\ \hline
O     & $0$     & $ \forall\,\, \omega $   & $\frac{3+3\omega-6n\omega}{n}$     \\ \hline
P     & $\frac{2n-1}{2n}$     & $n \neq 0$    & $\frac{-3+(6n-3)\omega}{n}$     \\ \hline
\end{tabular}
\caption{Fixed points with existence conditions and eigenvalues. (Model-I)}
\label{tablenew1}
\end{table}
\begin{table}[H]
\centering
\begin{tabular}{|p{2cm}|p{4cm}|p{5cm}|}
\hline
\textbf{Fixed Points} & \textbf{Stability} & \textbf{Cosmology} \\ \hline
O     &  stable for ($\omega >0  \land n> \frac{1+\omega}{2\omega}$), unstable for($\omega> 0  \land 0<n<\frac{1+\omega}{2\omega}$)      & Kasner-like with $a \sim t^{\frac{n}{3(1-(n-1)\omega)}}$   \\ \hline
P     & stable for ($\omega> 0  \land \frac{1}{2} \leq n<\frac{1+\omega}{2\omega}$), unstable for ($\omega> 0  \land n>\frac{1+\omega}{2\omega}$) & $a \sim t^{\frac{2n}{3(1+\omega)}}$    \\ \hline
\end{tabular}
\caption{{ Stability and cosmology of the fixed points. (Model-I)}\footnote{From this point, along the text we refer to the symbol $\land$ as AND, $\lor$ as OR.}}
\label{tablenew2}
\end{table}

We have noticed that this particular model has also been analysed in \cite{2, Esposito:2022omp}. In \cite{2}, the authors have obtained a 2-dimensional dynamical system for this model, whereas the authors in \cite{Esposito:2022omp} appear to have realised that the dynamical system for this simple case can be reduced to a 1-dimensional system. It is interesting to compare their results with what we obtained from the above analysis. 

In \cite{2}, the authors assume a very special case where $H_1=m H_2$ ($m=$constant), and obtains the fixed points, the associated eigenvalues and the effective equations of state $\omega_{\rm eff}$ numerically for particular chosen values of $m$. Therefore, their analysis can not be truly compared with ours. In \cite{Leon2024}, the authors have specialized particularly in a vacuum scenario, so they obtained fixed vacuum points. The effect of the inclusion of matter in our analysis is to introduce an additional fixed point $P$.

It is particularly interesting to compare the results from the analytical dynamical system analysis performed in \cite{Esposito:2022omp}, which adopts a covariant approach to study Bianchi type-I cosmological dynamics in $f(Q)$ gravity. The authors found that the phase space of $f(Q)=\alpha Q^n$ theory contained only one critical point $P_1$, which corresponds to our fixed point $P$. The cosmic evolution and the parameter range for (in)stability corresponding to this fixed point that we have obtained match that of $P_1$ from \cite{Esposito:2022omp}. However, we find an additional fixed point $O$, which, according to our analysis, does not violate any physical viability condition.

In passing, let us comment that we also correctly recover the known isotropic matter-dominated power law solution in the form of the fixed point $P$ (see \cite{R4}).

\subsection{Model-II: $f(Q) = Q + \alpha Q^{2}$}\label{model2}

The second example that we consider is the well-studied quadratic model. Wormhole solutions and cosmological perturbations in this model have been studied in \cite{52} and \cite{q2}, respectively. To find the viability of cosmological models, energy conditions have been explored using observational cosmographic data in \cite{60}. Nonsingular bouncing solutions in this model were studied in \cite{q3}. It is interesting to consider this model from the generic dynamical system formulation that we have provided.

For this model, we can express $Q$, $f_Q$, and $\Gamma(Q)$ in terms of the dynamical variables as follows. Solving $f_Q$ and $f_{QQ}$ for this specific model, one can obtain
\begin{align}
    f_Q=1+2\alpha Q, ~~ f_{QQ}=2\alpha.
    \label{fq1}
\end{align}
Explicit substitution of this specific model $f(Q)$ into (\ref{Qvar1}) gives
\begin{align}\label{Q1}
    Q = \frac{1}{\alpha} \left (\frac{2x_{1}+x_{3}-1}{3-4x_{1}-3x_{3}}\right),
\end{align}
putting (\ref{Q1}) into (\ref{fq1}), we get
\begin{align}\label{f_Q_example2}
f_{Q}= \frac{1-x_{3}}{3-4x_{1}-3x_{3}},
\end{align} 
Now substituting (\ref{Q1}) and (\ref{fq1}) into (\ref{Gamma}) we obtain,
\begin{align}
    \Gamma = \frac{1-x_{3}}{2(2x_{1}+x_{3}-1)}.
\end{align}
Substituting the above expression of $\Gamma$ into (\ref{gds1}), we have the following dynamical system,
\begin{align}\label{1MDSE1}
 x'_{1} = -3x_{1}(1+\omega)+6x_{1}(x_{1}+x_{3}+x_{1}\omega)+\frac{12x_{1}(x_{1}+x_{1}\omega)(2x_{1}+x_{3}-1)}{(x_{3}-1)(8x_{1}+3x_{3}-3)}, 
\end{align}
\begin{align} \label{1MDSE2}
 x'_{3} =6x_{3}x_{1}(1+\omega)+6x^{2}_{3}-6x_{3}. 
\end{align}

It is worth noting that the dynamical system is independent of the model parameter $\alpha$.

The variables $x_{1}$ and $x_{3}$ are nonnegative by definition. Combined with that, the physical viability condition $f_{Q}>0$ restricts the physically viable region of the phase space as follows,
\begin{align}
\left( 0 \leq x_{1} \leq \frac{3}{4} \right) \land 
\left( \left( 0 \leq x_{3} < \frac{1}{3} (3 - 4x_{1}) \right) \lor 
\left( x_{3} > 1 \right) \right) \lor 
\left( \left( x_{1} > \frac{3}{4} \right) \land 
\left( x_{3} > 1 \right) \right).
\label{1MC}
\end{align}
For this particular model \eqref{cosmology_new} gives
\begin{equation}
    \frac{\dot H}{H^2} = \frac{3 \left(3 (-1 + x_3)^2 x_3 + 8 x_1^2 x_3 (1 + \omega) + 
x_1 (-1 + x_3) \left(1 + \omega + x_3 (11 + 3 \omega)\right) \right)}
{(-1 + x_3)(-3 + 8 x_1 + 3 x_3)}.
\end{equation}
Based on the conditions, the physically viable region (\ref{1MC}) obtained for this quadratic model implies that, the only physically viable fixed points are O, P, and S, with S representing an anisotropic situation. The only saddle point indicating the matter-dominated phase is P but we have found that for $\omega=0 $ (dust), a matter-dominant cosmology e.g $a \sim t^{\frac{2}{3}}$ is not achieved. From the above-discussed results, we can infer that the quadratic model $f(Q)= Q+\alpha Q^{2}$ is not a cosmological model one at least for the late time cosmology and for the connection branch considered.\\
 The reference \cite{Rathore:2024ema} appears to have studied the stability aspect of this model in the context of an LRS Bianchi-I cosmology in $f(Q)$ theory. However, we do not quite agree with their definition of $\omega_{\rm eff}$ \cite[Eq.(42)]{Rathore:2024ema}, based on which they determine the cosmology corresponding to a fixed point. Therefore, we would not go into a comparison of our results with theirs.
\begin{table}[H]
\centering
\begin{tabular}{|p{2cm}|p{3.3cm}|p{2cm}|p{5.3cm}|}
\hline
\textbf{Fixed Points} & \textbf{Coordinates} & \textbf{Existence} & \textbf{Eigenvalues} \\ \hline
O     & (0,0)     &$ \forall\,\, \omega  $  & $-6, 3(-1-\omega)$     \\ \hline
P     & $(\frac{3-\sqrt{5}}{4},0)$     &  $\forall \,\,\omega  $   & $\frac{-9 (45 + 19 \sqrt{5}) (1 + \omega)}{2 (-3 + 2 \sqrt{5})^2}$, $\frac{-3}{2} \left(1 + \sqrt{5} - 3 \omega + \sqrt{5} \omega\right)$     \\ \hline
Q      & $(\frac{3+\sqrt{5}}{4},0)$       &  $ \forall\,\, \omega  $   & $\frac{9 (45 + 19 \sqrt{5}) (1 + \omega)}{2 (3 + 2 \sqrt{5})^2}$, $\frac{3}{2} \left(-1 + \sqrt{5} + 3 \omega + \sqrt{5} \omega\right)$       \\ \hline
R     & $(\frac{8}{5-4\omega+3\omega^{2}},\frac{3(-1-4\omega+\omega^{2})}{5-4\omega+3\omega^{2}})$      & $\omega< 2-\sqrt{5}$~ or $\omega > 2+\sqrt{5}$  & $\frac{3 (-45 + 21 \omega + 30 \omega^2 - 30 \omega^3 + 15 \omega^4 + 9 \omega^5 - \sqrt{3A})}{2 (1 + \omega) (-5 + 3 \omega) (5 - 4 \omega + 3 \omega^2)}$, $\frac{3 (-45 + 21 \omega + 30 \omega^2 - 30 \omega^3 + 15 \omega^4 + 9 \omega^5 + \sqrt{3A})}{2 (1 + \omega) (-5 + 3 \omega) (5 - 4 \omega + 3 \omega^2)}$    \\ \hline
S    & (0,1)     & $\forall\,\, \omega$     & $-6, -3(1+\omega)$     \\ \hline
\end{tabular}
\caption{ Fixed points with existence conditions and eigenvalues (Model-II). $A=1675 + 3170 \omega - 3153 \omega^2 + 456 \omega^3 + 4982 \omega^4 - 3972 \omega^5 - 10 \omega^6 + 2088 \omega^7 - 1473 \omega^8 + 306 \omega^9 + 27 \omega^{10}$.}
\label{table1}
\end{table}

\begin{table}[H]
\centering
\begin{tabular}{|p{2cm}|p{4cm}|p{5cm}|}
\hline
\textbf{Fixed Points} & \textbf{Stability} & \textbf{Cosmology} \\ \hline
O     &  stable for $\omega > -1$, non-hyperbolic for $\omega =-1$, saddle for $\omega < -1$      &  de Sitter ($H=constant$)    \\ \hline
P     & saddle for $-1< \omega < 2+\sqrt{5}$ &  $a \sim t^{\frac{4(3-2\sqrt{5})}{3(3-\sqrt{5}+\omega(3-\sqrt{5}))}}$    \\ \hline
Q      &  unphysical 
       & unphysical 
       \\ \hline
R     & unphysical & unphysical 

  \\ \hline
S     &  stable for $\omega > -1$, non-hyperbolic for $\omega =-1$, saddle for $\omega < -1$      & Kasner  \\ \hline
\end{tabular}
\caption{ Stability and cosmology of the fixed points.(Model-II)}
\label{table2}
\end{table}

\begin{figure}[H]
    \centering
    \includegraphics[width=0.4\textwidth]{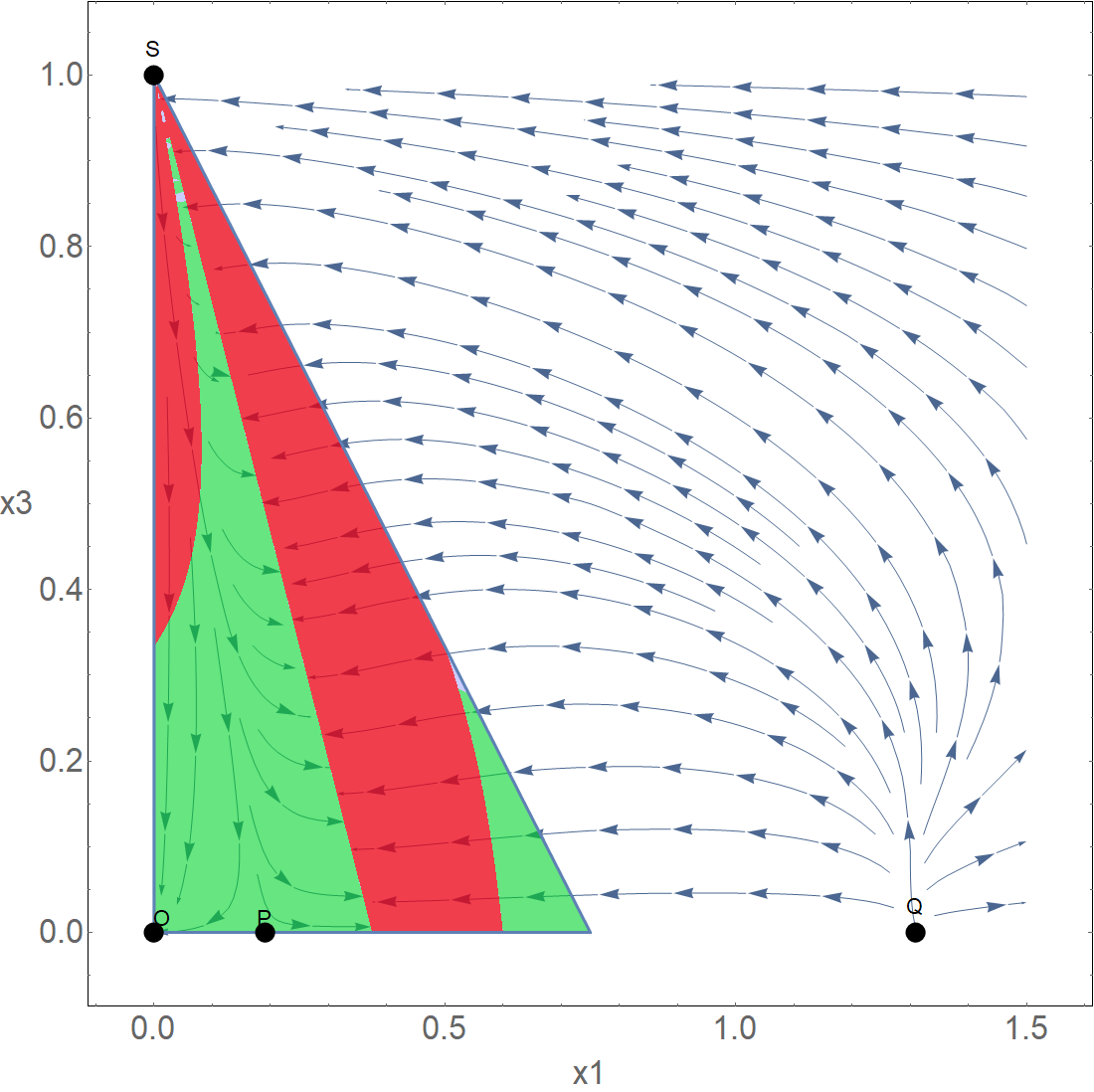}
    \caption{Phase Portrait of the model-II with $\omega =0$. Green region indicates the accelerated phase and the Red region indicates the decelerated phase of the universe.
   }
    \label{fig1}
\end{figure}

From Fig.\ref{fig1} one can see that, although the late time behaviour shows asymptotic isotropization, the only possible candidate for an intermediate matter-dominated epoch, namely the saddle fixed point $P$, falls within a region of the phase space that is characterised by an accelerated cosmology. This implies cosmological non-viability of the model $f(Q)=Q+\alpha Q^2$ in the context of late time cosmology. This conclusion does not depend on the sign or the value of the model parameter $\alpha$.



\subsection{Model-III: $f(Q) = Q + \alpha \sqrt {-Q} +\Lambda$}\label{model3}

This $f(Q)$ model has been studied in \cite{m2,Chakraborty:2025qlv}.  It indicates a real-valued Lagrangian $f(Q)$ that can, within the framework of the coincident gauge connection branch $\Gamma_1$, exclusively replicate the $\Lambda$CDM expansion history corresponding to a universe largely made up of dust-like matter.\footnote{Actually, it can be shown via the minisuperspace description of $f(Q)$ gravity that, at the level of the FLRW background and within the framework of the coincident gauge, this theory is equivalent to GR and therefore the scenario is equivalent to a General Relativistic $\Lambda$CDM scenario. This equivalent, however, will not hold at the level of the perturbed FLRW, and the theory will provide different signatures from the General Relativistic $\Lambda$CDM model \cite{Albuquerque:2022eac}.} By vanishing the $\Lambda$,  one can recover the dust-dominated universe other than GR. 

We can write the expressions for $Q$, $f_Q$, and $\Gamma(Q)$ in terms of the dynamical variables in the following way. We can write $f_Q$ and $f_{QQ}$ for this particular model as 
\begin{align}
    f_Q=1-\frac{\alpha}{2\sqrt{-Q}}, ~~ f_{QQ}=-\frac{\alpha}{4(-Q)^\frac{3}{2}}.
    \label{fq2}
\end{align}
Explicit substitution into (\ref{Qvar1}) gives
\begin{align}
    \sqrt{-Q} = \frac{\alpha x_{1} \pm \sqrt{\alpha^{2} x_{1}^{2}-4(x_{3}-1+2x_{1})(\Lambda x_{3}-\Lambda)}}{2(x_{3}+2x_{1}-1)}.
    \label{negQ}
\end{align}
Putting (\ref{negQ}) into (\ref{fq2}), one can get
\begin{align}
f_{Q}=\frac{(1-x_{3}-x_{1})\pm \sqrt{x_{1}^{2}-4c(x_{3}+2x_{1}-1)(x_{3}-1)}}{x_{1} \pm \sqrt{x_{1}^{2}-4c(x_{3}+2x_{1}-1)(x_{3}-1)}}, \qquad \left(c=\frac{\Lambda}{\alpha^{2}}\right)
\end{align}
Utilizing (\ref{fq2}) and (\ref{negQ}) in (\ref{Gamma}), we have
\begin{align}
   \Gamma = \frac{2 x_{1} \pm 2\sqrt{x_{1}^{2}-4c(x_{3}-1+2x_{1})( x_{3}-1)}}{ (x_{3}+2x_{1}-1)} -2 .
\end{align}
For $\Gamma$ to be real-valued, $c$ must be in the following range i.e,
\begin{align}
   -\frac{1}{4}\leq c \leq 0.
   \label{condition:c}
\end{align}
Since $c=\frac{\Lambda}{\alpha^{2}}$, the above condition implies that $\Lambda$ must be non-positive i.e $\Lambda \leq 0$. The dynamical system equations for this model are,
\begin{align}
x'_{1} = -3x_{1}(1+\omega) +6x_{1}(x_{1}+x_{3}+x_{1}\omega)+ \frac{6x_{1}^2 (1+\omega)(x_{3}+2x_{1}-1)}{(x_{3}-1)\left(2x_{1} \pm 2 \sqrt{x_{1}^{2}-4c(x_{3}-1+2x_{1})(x_{3}-1)}\right)},
\end{align}
\begin{align}
x'_{3} = 6x_{3}x_{1}(1+\omega)+6x^{2}_{3}-6x_{3}.
\end{align}

\begin{figure}[htbp]
    \centering
    \begin{subfigure}[b]{0.45\textwidth}
        \includegraphics[width=\linewidth]{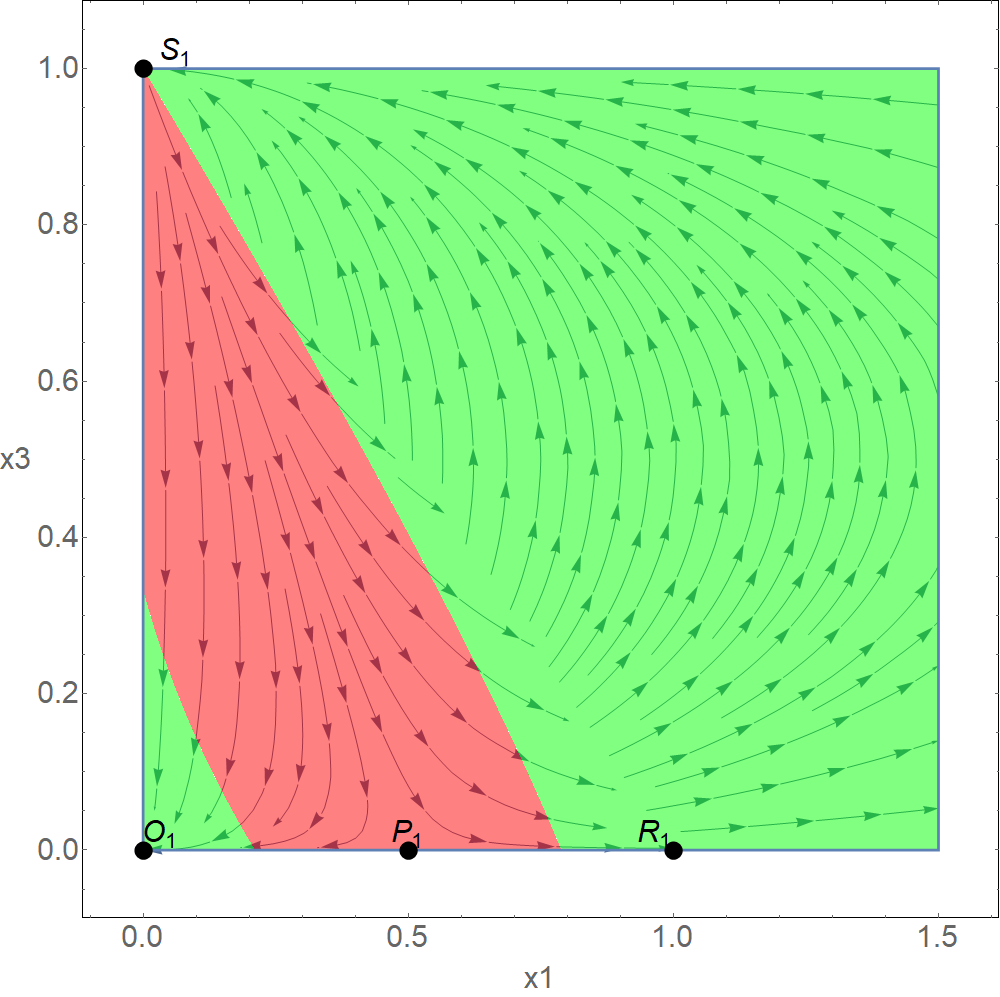}
        \caption{Phase Portrait of the model-III for $\omega = 0$ and $c = -\frac{1}{4}$. Green region indicates the accelerated phase and red region indicates the decelerated phase of the universe.}
        \label{fig2}
    \end{subfigure}
    \hfill
    \begin{subfigure}[b]{0.45\textwidth}
        \includegraphics[width=\linewidth]{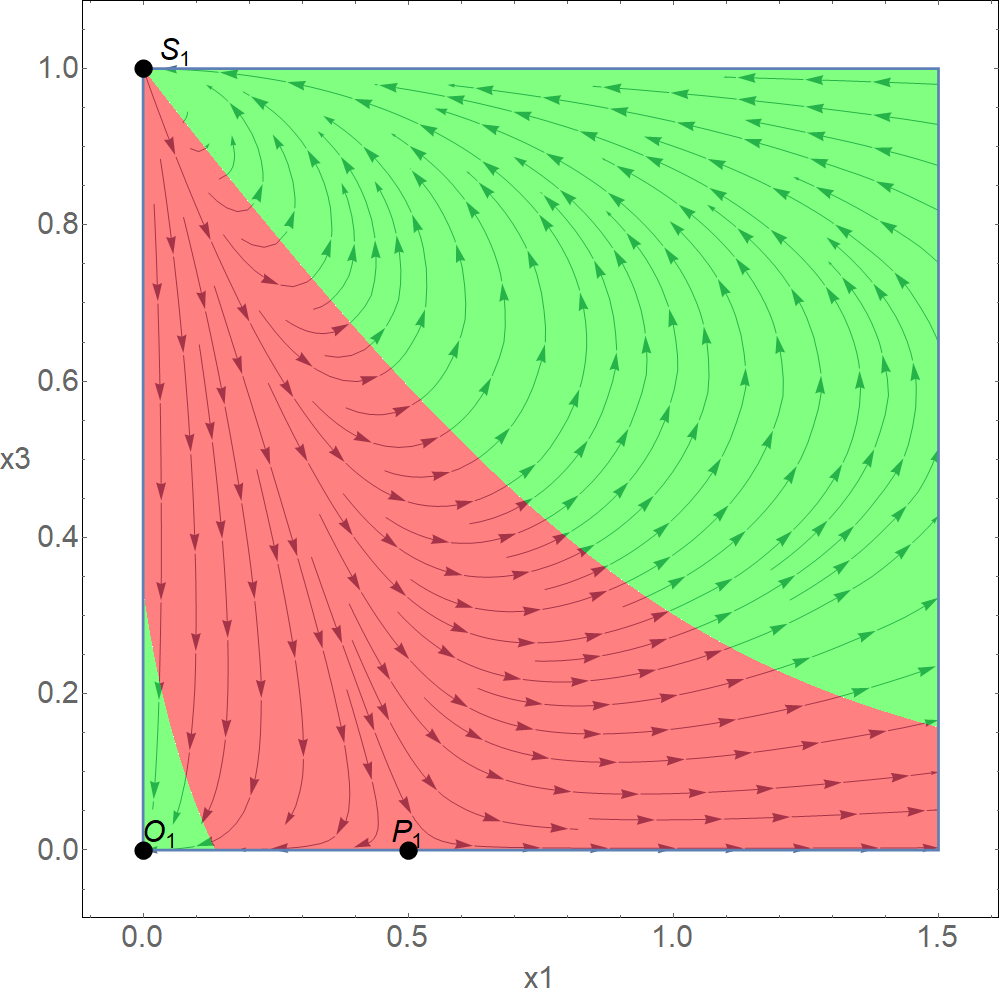}
        \caption{Phase Portrait of the model-III for $\omega = 0$ and $c = -\frac{1}{25}$. Green region indicates the accelerated phase and red region indicates the decelerated phase of the universe.}
        \label{fig3}
    \end{subfigure}
    \caption{Comparison of cosmological phase portraits for different values of $c$ in model-III.}
    \label{fig:sidebyside}
\end{figure}

Although there are two model parameters $\alpha,\Lambda$, the dynamical system for this model depends only on the combination $c=\frac{\Lambda}{\alpha^2}$. 
For this particular model \eqref{cosmology_new} gives
{\scriptsize
\begin{align}
    \frac{\dot{H}}{H^2}=3\left(\frac{\left(2x_3(x_3+2x_1-1)+(2x_1\pm2\sqrt{x_1^2-4c(x_3-1+2x_1)(x_3-1)}-2(x_3+2x_1-1)(x_3-1))\right)(-x_3-x_1-x_1\omega)+2x_3(x_3+2x_1-1)}{\left(2x_1\pm2\sqrt{x_1^2-4c(x_3-1+2x_1)(x_3-1)}\right)(x_3-1)}\right).
\end{align}
}
The variables $x_{1}$ and $x_{3}$ are nonnegative by definition. Combined with that, the physical viability condition $f_{Q}>0$ restricts the physically viable region of the phase space as follows,
\begin{equation}
 (x_{1} \geq 0) \land (0 \leq x_{3} < 1).
\label{2MVR}
\end{equation}
In Table \ref{table3} and Table \ref{table4}, one can find the existence conditions for the critical points, eigenvalues, stability conditions, and cosmology for this model. 
\begin{table}[H]
\centering
\begin{tabular}{|p{0.7cm}|p{1.7cm}|p{1.5cm}| p{11cm}|}
\hline
\textbf{Fixed Points} & \textbf{Coordinates} & \textbf{Existence} & \textbf{Eigenvalues} \\ \hline
$ O_{1}$     & (0,0)     & $ \forall ~ \omega$ and $c$     & $-6, 3(-1-\omega)$     \\ \hline
$P_{1}$     & $(\frac{1}{2},0)$     & $ \forall  ~ \omega$ and $c$      & $3 (-1 + \omega), \frac{3 (1+ \omega)}{2}$  \\ \hline
$R_{1}$     & $(2(-2 c+\sqrt{c +4 c^{2} }),0)$  & $c \leq -\frac{1}{4}$ or $ c > 0$  & $3 \left( 1+\omega \right) \sqrt{ c \left( 1+4c - 2\sqrt{ c \left( 1+4c \right) } \right) } \left( \frac{1}{\sqrt{2c \left( 1+4c \right)}} +  \frac{\sqrt{2}(2\sqrt{ c \left( 1+4c \right) } - 1)}{1+4c} - \frac{4\sqrt{2}c}{1+4c} \right),  
\newline
6 \left( -1 + 2\sqrt{c(1+4c)} + 2\sqrt{c(1+4c)}\omega - 4c(1+\omega) \right)$ \\ \hline
$S_{1}$     & (0,1)     & $ \forall ~ \omega$ and $c$      & $-6, -3(1+\omega)$     \\ \hline
\end{tabular}
\caption{ Fixed points with existence conditions and eigenvalues. (Model-III)}
\label{table3}
\end{table}
\begin{table}[H]
\centering
\begin{tabular}{|p{2cm}|p{4cm}|p{5cm}|}
\hline
\textbf{Fixed Points} & \textbf{Stability} & \textbf{Cosmology} \\ \hline
$O_{1}$     &   stable for $\omega > -1$, non-hyperbolic for $\omega =-1$, saddle for $\omega < -1$ &  de Sitter ($H=constant$)  \\ \hline
$P_{1}$     &  saddle for $-1< \omega < 1$ &  $a \sim t^{\frac{2}{3(1+\omega)}}$     \\ \hline
$R_{1}$     &  saddle for $ c>0$ and $-1<\omega \leq 1$&$ a \sim t^{\frac{1}{3\left((1+4c - 2\sqrt{c + 4c^2})(1 + \omega)\right)}} $ \\ \hline
$S_{1}$     &   stable for $\omega > -1$, non-hyperbolic for $\omega =-1$, saddle for $\omega < -1$ & Kasner \\ \hline
\end{tabular}
\caption{ Stability and cosmology of the fixed points. (Model-III)}
\label{table4}
\end{table}
We have added phase portraits in the presence of dust ($\omega=0$) for two values of $c$ that lie within the range of $c$ as mentioned in (\ref{condition:c}). We notice that for $c=-\frac{1}{4}$, the critical point $R_{1}$ falls within the physical viable region of the real phase portrait as given in the Fig \ref{fig2}, and for $c=-\frac{1}{25}$ it doesn't exist within the physical viable region as provided in Fig \ref{fig3}. The saddle fixed point $P_1$ serves as the intermediate matter-dominated epoch, providing the cosmological evolution $a \sim t^{\frac{2}{3}}$, whereas the future attractor $O_1$ serves as the dark energy dominated epoch. 

From the figures \ref{fig2} and \ref{fig3} it is clear that there exists asymptotic isotropization in late time cosmology in the model $f(Q)=Q+\alpha\sqrt{-Q}+\Lambda$ as long as $-\frac{\alpha^2}{4}\leq\Lambda\leq0$ \eqref{condition:c}. In other words, the model is stable against small homogeneous anisotropic perturbations in the spacetime geometry. This implies that this model can serve as a viable candidate for late-time modified gravity. This conclusion does not depend on the sign or the value of the model parameter $\alpha$. 

\subsection{Model-IV: $f(Q) = Q e^ {\lambda \frac{Q_{0}}{Q}}$}\label{model4}

This power exponential model has been studied in \cite{55}, it has passed the constraints of the early times. Moreover, different observational data have been analyzed against this model, and it has been established that this model is comparable for some data sets, while for others, it is preferable to $\Lambda$CDM. $\lambda$ is the free parameter of the model, we get GR without a cosmological constant if $\lambda$ vanishes. But it has no $\Lambda$CDM limit. With small $\lambda$, it reduces to GR with $Q_{0}\lambda$ as a cosmological term. The forms of $Q$, $f_Q$, and $\Gamma$ in terms of variables for our study can be obtained as: we have
\begin{align}
    f_Q=e^{\frac{\lambda Q_0}{Q}}\left(1-\frac{\lambda Q_0}{Q}\right), ~~ f_{QQ}=e^{\frac{\lambda Q_0}{Q}}\left(\frac{\lambda^2 Q_0^2}{Q^3}\right).
    \label{fq3}
\end{align}
Explicit substitution into (\ref{Qvar1}) gives
\begin{align}
    Q = \frac{2x_{2} \lambda Q_{0}}{2x_{2}-1+x_{3}},
    \label{Qe}
\end{align}
substituting  (\ref{Qe}) into (\ref{fq3}), one can get
\begin{align}
f_Q = e^{\frac{1 - x_3 - 2 x_1}{2 - 2 x_1 - 2 x_3}} \left( \frac{1 - x_3}{2 - 2 x_1 - 2 x_3} \right),
\end{align}
using (\ref{fq3}) and (\ref{Qe}) into (\ref{Gamma}), one can get
\begin{align}
   \Gamma = \frac{4(1-x_{1}-x_{3})^{2}}{(1-2x_{1}-x_{3})^{2}}-\frac{2 (1-x_{1}-x_{3})}{1-2x_{1}-x_{3}}.
\end{align}
Replacing this value of $\Gamma$ in (\ref{gds1}), we get the dynamical system of this model as, 
\begin{align} \label{3MDSE1}
x'_1 &= -3x_1(1+\omega) + 6x_1(x_1 + x_3 + x_1\omega) \nonumber \\
&\quad + \frac{3x_1(x_1 + x_1\omega)(1 - 2x_1 - x_3)^2}
{(x_3 - 1)\left[2(1 - x_1 - x_3)^2 - (1 - x_1 - x_3)(1 - 2x_1 - x_3) + (1 - 2x_1 - x_3)^2\right]},
\end{align}
\begin{align}\label{3MDSE2}
x'_3 &= 6x_3x_1(1+\omega)+6x^{2}_{3}-6x_{3}. 
\end{align}

For this example \eqref{cosmology_new} gives
{\scriptsize
\begin{align}
    \frac{\dot{H}}{H^2}=3\left(\frac{\left(2x_3(1-2x_1-x_3)^2 +\left(4(1-x_1-x_3)^2-2(1-x_1-x_3)(1-2x_1-x_3)\right)(x_3-1))\right)(-x_3-x_1-x_1\omega)+2x_3(1-2x_1-x_3)^2}{\left(4(1-x_1-x_3)^2-2(1-x_1-x_3)(1-2x_1-x_3)+2(1-2x_1-x_3)^2\right)(x_3-1)}\right).
\end{align}
}
It is evident that the dynamical system of this model is independent of both the model parameters $Q_0$ and $\lambda$.

The variables $x_{1}$ and $x_{3}$ are nonnegative by definition. Combined with that, the physical viability condition $f_{Q}>0$ restricts the physically viable region of the phase space as follows
\begin{align}
\Bigg(\left( 0 \leq x_{1} < 1 \right) \land \left( 0 \leq x_{3} < 1 - x_{1} \right) \Bigg) \lor 
\Bigg(\left( x_{1} > 0 \right) \land \left( x_{3} > 1 \right) \Bigg).
\label{3MVR}
\end{align}

\begin{figure}[H]
    \centering
    \includegraphics[width=0.4\textwidth]{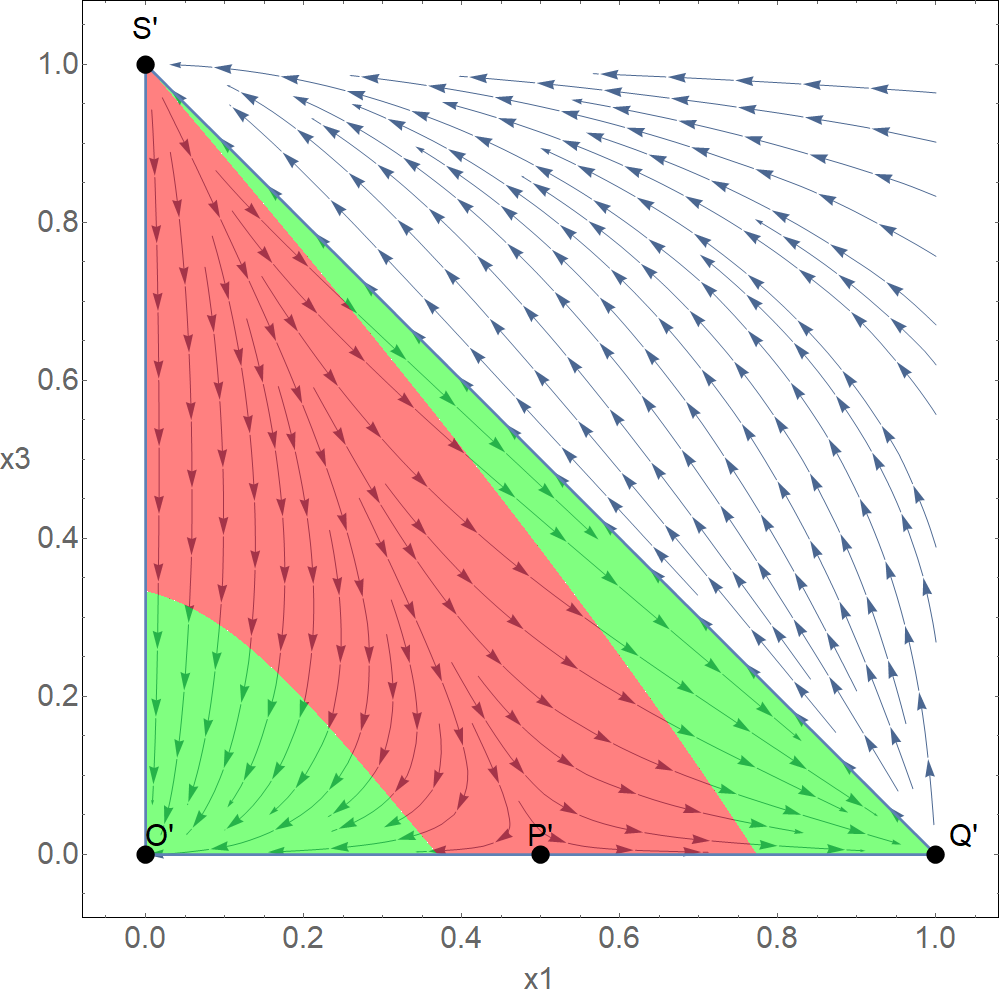}
    \caption{Phase Portrait for the model-IV with $\omega =0$. Green region indicates the accelerated phase and Red region indicates the decelerated phase of the universe.}
    \label{fig4}
\end{figure}

The fixed points, the conditions under which they exist, and the eigenvalues corresponding to each critical point of the jacobian matrix are presented in Table \ref{table5}. The stability conditions and cosmology are provided in Table \ref{table6}. 
\begin{table}[H]
\centering
\begin{tabular}{|p{0.7cm}|p{3.8cm}|p{1.5cm}| p{4cm}|}
\hline
\textbf{Fixed Points} & \textbf{Coordinates} & \textbf{Existence} & \textbf{Eigenvalues} \\ \hline
$O^{'}$     & (0,0)     & $ \forall ~\omega$     & $-6, 3 (-1 - \omega)$   \\ \hline
$P^{'}$     & $(\frac{1}{2},0)$     & $\forall ~  \omega$     & $3 (-1 + \omega), 3 (1 + \omega)$    \\ \hline
$Q^{'}$      & $(1,0)$        & $\forall ~ \omega$    & $6\omega, 0$   \\ \hline
$S^{'}$     & (0,1)     & $\forall~ \omega$      & $-6, -3 (1 + \omega)$   \\ \hline
\end{tabular}
\caption{ Fixed points with existence conditions and eigenvalues. (Model-IV)}
\label{table5}
\end{table}
\begin{table}[H]
\centering
\begin{tabular}{|p{2cm}|p{4cm}|p{4cm}|}
\hline
\textbf{Fixed Points} & \textbf{Stability} & \textbf{Cosmology} \\ \hline
$O^{'}$     &   stable for $\omega > -1$, non-hyperbolic for $\omega =-1$, saddle for $\omega < -1$ & de Sitter ($H=constant$)\\ \hline
$P^{'}$    &   saddle for $-1< \omega < 1$ , &  $a \sim t^{\frac{2}{3(1+\omega)}}$   \\ \hline
$Q^{'}$   &  non-hyperbolic for  $\omega=0$ & de Sitter ($H=constant$)\\ \hline
$S^{'}$     &   stable for $\omega > -1$, non-hyperbolic for $\omega =-1$, saddle for $\omega < -1$ & Kasner  \\ \hline
\end{tabular}
\caption{ Stability and cosmology of the fixed points. (Model-IV) }
\label{table6}
\end{table}
The phase portrait of the model in the presence of dust ($\omega=0$) is presented in Fig.\ref{fig4}. The saddle fixed point $P^{'}$ serves as the intermediate matter-dominated epoch, providing the cosmological evolution $a \sim t^{\frac{2}{3}}$, whereas the future attractor $O'$ serves as the dark energy dominated epoch.
From the figure \ref{fig4} it is clear that there exists asymptotic isotropization in late time cosmology in the model $f(Q)=Q e^ {\lambda \frac{Q_{0}}{Q}}$. In other words, the model is stable against small homogeneous anisotropic perturbations in the spacetime geometry. Moreover, this conclusion does not depend on the sign or the value of any of the model parameters $Q_0$ or $\lambda$. This makes the model particularly attractive as a viable candidate for late-time modified $f(Q)$ gravity. 

\section{General formulation of the Dynamical system for $\Gamma_{II}$}\label{dsaII}

In this section, we present the similar generic construction methodology of the dynamical system for any given $f(Q)$ gravity model corresponding to the connection class $\Gamma_2$. We derive the non-metricity scalar $Q$ and the Friedmann equations corresponding to this connection class \footnote{Upon re-deriving the connection field equations and the pressure equation, we identified some typographical inconsistencies in the original presentation given in \cite{Leon2024}. Our corrected version aligns with the standard equations in FLRW ($\sigma=0$) for the same connection class \cite{80}.}
\begin{equation}\label{Q2}
    Q=-6H^2+\sigma^2+9H\gamma+3\dot{\gamma}.
\end{equation}
\begin{align}
   \kappa \rho=\frac{1}{2}f+(3H^2-\frac{Q}{2}-\frac{\sigma^2}{2})f_{Q}+\frac{3}{2}\gamma\dot{Q}f_{QQ},
\end{align}

\begin{align}\label{secp}
   \kappa p=-\frac{1}{2}f+(-2\dot{H}-3H^2+\frac{Q}{2})f_{Q}+\frac{\dot{Q}}{2}(-4H+3\gamma)f_{QQ},
\end{align}
We obtain the same anisotropy evolution equation as in the first connection class (see Appendix \ref{dotsigma})
\begin{align}\label{connllsigma}
     \dot{\sigma}= -\sigma\left(\frac{\dot{f}_{Q}}{f_{Q}}+3H\right),
\end{align}
which can be integrated as
\begin{align}
    \sigma \sim \frac{1}{a^3f_Q}.
\end{align}
The connection field equation is
\begin{equation}\label{connIIeq}
    \gamma\left(\ddot f_Q+3H\dot f_Q\right)=0,
\end{equation}
which can be integrated to give (since $\gamma$ is by definition nonzero)
\begin{equation}
    \dot{f}_Q = f_{QQ}\dot{Q} \sim \frac{1}{a^3}.
\end{equation}
As usual, in the absence of hypermomentum, the standard continuity relation holds
\begin{align}
    \dot \rho +3H(p+\rho)=0.
\end{align}
We define the dimensionless variables
$$x_1=-\frac{f}{6H^2f_Q},~x_2=\frac{Q}{6H^2}~,x_3=-\frac{\dot{Q}f_{QQ}}{Hf_Q},~x_4=\frac{\gamma}{2H},~x_5=\frac{\kappa \rho}{3H^2f_Q},~x_6=\frac{\sigma^2}{6H^2},~\Gamma=\frac{f_Q}{Qf_{QQ}}.$$
The constraint eq is 
\begin{align}\label{constr2}
    x_1=1-x_2-x_3x_4-x_5-x_6,
\end{align}
We can also write
\begin{align}\label{29a}
\frac{Qf_Q}{f}=\frac{x_2}{x_2+x_3x_4+x_5+x_6-1},
\end{align}
and also we have
\begin{align}\label{dotH}
    \frac{\dot{H}}{H^2}=(1-3x_4)x_3-\frac{3}{2}\left(x_5(1+\omega)+x_6\right).
\end{align}
A noteworthy point here is that, whereas the expression for $\frac{\dot{H}}{H^2}$ was dependent on the quantity $\Gamma\equiv\frac{f_Q}{Q f_{QQ}}$, i.e. on the particular $f(Q)$ model for the connection class $\Gamma_1$ (see Eq.\eqref{cosmology_new}), it is model-independent for the connection class $\Gamma_2$.

Now using the eq (\ref{Q2}), (\ref{secp}), (\ref{connllsigma}), (\ref{connIIeq}), (\ref{conteq}), (\ref{constr2}) and (\ref{dotH}). The autonomous system is given as
\begin{align}
    x_2'=-x_2 x_3 \Gamma -2x_2 \left ((1-3x_4)x_3-\frac{3}{2}(x_5(1+\omega)+x_6)\right),
\end{align}
\begin{align}
    x_3'=-3x_3+x_3^2-x_3\left(x_3(1-3x_4)-\frac{3}{2}x_5(1+\omega)-\frac{3}{2}x_6\right),
\end{align}
\begin{align}
    x_4'=1+x_2-3x_4-x_6-x_4 \left(x_3-3x_3x_4 -\frac{3}{2}x_5 (1+\omega)-\frac{3}{2}x_6\right),
\end{align}
\begin{align}
    x_5'=-3x_5 (1+\omega)+x_5 x_3 -2x_5 \left(x_3(1-3x_4)-\frac{3}{2}x_5(1+\omega)-\frac{3}{2}x_6\right),
\end{align}
\begin{align}\label{x6'}
    x_6'=2x_3x_6-6x_6-2x_6 \left(x_3-3x_3x_4 -\frac{3}{2}x_5 (1+\omega)-\frac{3}{2}x_6\right).
\end{align}


\section{Application in the context of late-time cosmology for $\Gamma_{II}$}\label{lateG2}

We proceed in the same manner as in section \ref{sec5} to demonstrate the applicability of our generic dynamical system formulation for the connection branch $\Gamma_{II}$ by applying it explicitly to the same four $f(Q)$ models that we have considered for the connection branch $\Gamma_I$. We write the generic dynamical equations for each of the models, but specialise to the case of dust ($\omega=0$) for the fixed point analysis.

\subsection{Model-I: $f(Q) = \alpha(-Q)^n$ ($n\neq1$)}\label{model1-conll}

As a first example, we consider the simplest power law model with constant $\alpha$ and $n$. Now solving $f_Q$ and $f_{QQ}$ for this specific model, one can obtain
\begin{align}
    f_Q=-n\alpha(-Q)^{n-1}, ~~ f_{QQ}=n(n-1)\alpha(-Q)^{n-2}.
    \label{fq_cll}
\end{align}
Explicit substitution of this specific model $f(Q)$ into (\ref{29a}) gives us an additional constraint,
\begin{align}\label{anotherconstr_cll}
    x_3=\frac{n(1-x_5-x_6)+(1-n)x_2)}{nx_4},
\end{align}
one can also rewrite the $f_Q$ as
\begin{align}\label{fqeq_const}
    f_Q=n\alpha (-1)^n (6H^2x_2)^{n-1}.
\end{align}
Substituting (\ref{fq_cll}) into (\ref{Gamma}) we obtain,
\begin{align}
    \Gamma = \frac{1}{n-1}.
\end{align}
Under this setting, the dynamical system is written as,
\begin{align}
    x_2'=-\frac{x_2\left(n(1-x_5-x_6)+(1-n)x_2)\right)}{n(n-1)x_4} -2x_2 \left (\frac{(1-3x_4)\left( n(1-x_5-x_6)+x_2(1-n)\right)}{nx_4}-\frac{3\left(x_5(1+\omega)+x_6\right)}{2}\right),
\end{align}
\begin{align}
    x_4'=1+x_2-3x_4-x_6-x_4 \left(
    \frac{(1-3x_4)\left( n(1-x_5-x_6)+x_2(1-n)\right)}{nx_4}-\frac{3\left(x_5(1+\omega)+x_6\right)}{2}\right),
\end{align}
\begin{align}
    x_5'=\,& -3x_5 (1+\omega)+x_5 \left(\frac{n(1-x_5-x_6)+(1-n)x_2)}{nx_4}\right)  \nonumber \\
& -2x_5 \left(\frac{(1-3x_4)\left( n(1-x_5-x_6)+x_2(1-n)\right)}{nx_4}-\frac{3\left(x_5(1+\omega)+x_6\right)}{2}\right),
\end{align}
\begin{align}
    x_6'=2x_6\left(\frac{n(1-x_5-x_6)+(1-n)x_2)}{nx_4}\right)-6x_6-2x_6 \left(\frac{(1-3x_4)\left( n(1-x_5-x_6)+x_2(1-n)\right)}{nx_4}-\frac{3\left(x_5(1+\omega)+x_6\right)}{2}\right).
\end{align}
For this particular model \eqref{dotH} gives
\begin{equation}
    \frac{\dot H}{H^2} =\frac{(1-3x_4)\left( n(1-x_5-x_6)+x_2(1-n)\right)}{nx_4}-\frac{3\left(x_5(1+\omega)+x_6\right)}{2}.
\end{equation}

\begin{table}[H]
\centering
\begin{tabular}{|p{2cm}|p{4cm}|p{4cm}|}
\hline
\textbf{Fixed Points} & \textbf{$(x_2,x_4,x_5,x_6)$} & \textbf{Existence}  \\ \hline
$A_1$     & $(x_2,\frac{2}{3}(1-x_2),0,2x_2)$       & unphysical \\ \hline
$A_2$    & $(0,\frac{1}{3},0,0)$      &unphysical\\ \hline
$A_3$     &  $(0,\frac{2}{3},1,0)$    & unphysical  \\ \hline
$A_4$    &  $(\frac{n}{3(n-2)},\frac{4}{9},\frac{1+2n}{3(2-n)},0)$   &unphysical \\ \hline
$A_5$    &  $(0,\frac{2n-3}{6(n-1)},0,0)$   &unphysical  \\ \hline
$A_{6}$  &  $(\frac{n}{n-1},\frac{3-2n}{6(1-n)},0,0)$   & for even n: $ n\neq 1 \land \alpha>0\land n>1$ ,  for odd n: $ n\neq 1 \land \alpha<0\land n>1$\\ \hline

\end{tabular}
\caption{Critical points and their existence conditions for non-relativistic matter($\omega=0$) (Model-I).}
\label{table1_cll}
\end{table}


\begin{table}[H]
\centering
\begin{tabular}{|p{2cm}|p{4cm}|p{2cm}|p{4cm}|}
\hline
\textbf{Fixed Points} & \textbf{Eigenvalues} & \textbf{Stability} &\textbf{Cosmology}  \\ \hline
$A_6$     & $(0,-6,-3,-3) $    &stable &  de Sitter ($H=constant$) \\ \hline
\end{tabular}
\caption{Eigenvalues, stability, and cosmology of the viable fixed points (Model-I). }
\label{table2_cll}
\end{table}
The detailed analysis of the fixed points is presented in Table \ref{table1_cll} and Table \ref{table2_cll}. In this power-law model, five critical points associated with isotropic FLRW cosmologies are identified, along with a non-isolated 1-parameter family of anisotropic solutions. However, when subject to the viability conditions \{$x_5\geq 0,~x_6\ge 0$ and $f_Q>0$\} (the condition $f_Q>0$ can be investigated by substituting the coordinates of the fixed point into the expression (\ref{fqeq_const})), it turns out that all fixed points, including the anisotropic family, fail to satisfy this viability criterion. As a result, all the fixed points except $A_6$ are rendered unphysical in the context of this theory. The fixed point $A_6$ describes a stable de-Sitter future attractor. 

We note that the Bianchi-I phase space of this $f(Q)$ model in the absence of matter has also been studied in the connection branch $\Gamma_2$ in \cite{Leon2024}, where the author identified three families of critical points; one isotropic de-Sitter and two anisotropic Kasner-like. The inclusion of matter into the picture introduces additional fixed points in the phase space e.g. $A_3$ and $A_4$. The families Kasner-like fixed points obtained in \cite{Leon2024} match with our family of fixed points $A_1$, and the family of stable de-Sitter attractor obtained in \cite{Leon2024} matches with our fixed point $A_6$. The isotropic de Sitter attractor was also found in the phase space analysis of FLRW solutions in Ref.\cite{R4}. This hints towards the consistency of our findings with the earlier works. However, our analysis reveals the existence of additional isotropic fixed points in the phase space, e.g. $A_2$ and $A_5$. Moreover, one key improvement provided by our analysis is the clear emphasis on singling out the physically viable region, which clearly shows that the Kasner-like solutions fall into a region of the phase space that violates the physical viability condition $f_Q>0$.

\subsection{Model-II: $f(Q) = Q + \alpha Q^{2}$}
For this example, we can express $Q$, $f_Q$, and $\Gamma(Q)$ in terms of the dynamical variables as follows. Solving $f_Q$ and $f_{QQ}$ for this specific model, one can obtain
\begin{align}
    f_Q=1+2\alpha Q, ~~ f_{QQ}=2\alpha.
    \label{fqq}
\end{align}
Explicit substitution of this specific model $f(Q)$ into (\ref{29a}) gives
\begin{align}\label{QQ}
    Q = \frac{1}{\alpha} \left (\frac{1-x_3x_4-x_5-x_6}{x_2+2x_3x_4+2x_5+2x_6-2}\right),
\end{align}
putting (\ref{QQ}) into (\ref{fqq}), we get
\begin{align}\label{}
f_{Q}= \frac{x_2}{x_2+2x_3x_4+2x_5+2x_6-2},
\end{align} 
Now substituting (\ref{fqq}) and (\ref{QQ}) into (\ref{Gamma}) we obtain,
\begin{align}
    \Gamma = \frac{x_2}{2(1-x_3x_4-x_5-x_6)}.
\end{align}
using all, we have the following autonomous dynamical system,
\begin{align}
    x_2'=-\frac{x_2^2x_3}{2(1-x_3x_4-x_5-x_6)} -2x_2 \left ((1-3x_4)x_3-\frac{3}{2}(x_5(1+\omega)+x_6)\right),
\end{align}
\begin{align}
    x_3'=-3x_3+x_3^2-x_3\left(x_3(1-3x_4)-\frac{3}{2}x_5(1+\omega)-\frac{3}{2}x_6\right),
\end{align}
\begin{align}
    x_4'=1+x_2-3x_4-x_6-x_4 \left(x_3-3x_3x_4 -\frac{3}{2}x_5 (1+\omega)-\frac{3}{2}x_6\right),
\end{align}
\begin{align}
    x_5'=-3x_5 (1+\omega)+x_5 x_3 -2x_5 \left(x_3(1-3x_4)-\frac{3}{2}x_5(1+\omega)-\frac{3}{2}x_6\right),
\end{align}
\begin{align}
    x_6'=2x_3x_6-6x_6-2x_6 \left(x_3-3x_3x_4 -\frac{3}{2}x_5 (1+\omega)-\frac{3}{2}x_6\right).
\end{align}
The variables $x_{5}$ and $x_{6}$ are nonnegative by definition. Combined with that, the physical viability condition $f_{Q}>0$ restricts the physically viable region of this model as follows,
\begin{align}\label{QMvc}
&\Big( 0 \leq x_6 \leq 1 \land  0 \leq x_5 \leq 1 \land  x_2 \leq 0  \land  
 x_4 > 0  \land x_3 < \frac{2 - x_2 - 2x_5 - 2x_6}{2x_4}
\Big).
\end{align}

\begin{table}[H]
\centering
\begin{tabular}{|p{2cm}|p{5cm}|p{4cm}|}
\hline
\textbf{Fixed Points} & \textbf{$(x_2,x_3,x_4,x_5,x_6)$} & \textbf{Existence}  \\ \hline
$P_1$     & $(x_2,0,\frac{1+x_2}{3},0,0) $    &$ -1<x_2 \leq0 ~\land 0<x_4 \leq \frac{1}{3} $    \\ \hline
$P_2$    & $(0,3,x_4,1-3x_4,1-3x_4)$      &unphysical \\ \hline
$P_3$    & $(x_2,4,\frac{1-x_2}{4},0,2x_2)$     & unphysical    \\ \hline
\end{tabular}
\caption{Critical points and their existence conditions for non-relativistic matter($\omega=0$) (Model-II).}
\label{table3_cll}
\end{table}


\begin{table}[H]
\centering
\begin{tabular}{|p{2cm}|p{5cm}|p{1.5cm}|p{4cm}|}
\hline
\textbf{Fixed Points} & \textbf{Eigenvalues} & \textbf{Stability}&\textbf{Cosmology}  \\ \hline
$P_1$     & $(-6,-3,-3,-3,0) $    &stable & de Sitter ($H=constant$)   \\ \hline
\end{tabular}
\caption{Eigenvalues, stability, and cosmology of the viable fixed points (Model-II).}
\label{table4_cll}
\end{table}
For the quadratic model, a brief description of the stationary points is provided in Table \ref{table3_cll} and Table \ref{table4_cll}. The critical point $P_1$ corresponds to a family of isotropic solutions and represents a future de Sitter universe. However, the anisotropic families of critical points $P_2$ and $P_3$ do not satisfy the physically viable region defined by the conditions in (\ref{QMvc}), thereby rendering them unphysical within the context of this model.

\subsection{Model-III: $f(Q) = Q + \alpha \sqrt {-Q} +\Lambda$}\label{model3_connll}
We can write $f_Q$ and $f_{QQ}$ for this particular model as 
\begin{align}
    f_Q=1-\frac{\alpha}{2\sqrt{-Q}}, ~~ f_{QQ}=-\frac{\alpha}{4(-Q)^\frac{3}{2}}.
    \label{fq2_cll}
\end{align}
Explicit substitution into (\ref{29a}) gives
\begin{align}
    \sqrt{-Q} = \frac{\alpha(-1+x_2(x_3-1)+x_5+x_6)) \pm \sqrt{\alpha^{2} (-1+x_2(x_3-1)+x_5+x_6)^2-16\Lambda x_3(-1+x_3x_4+x_5+x_6)}}{4(x_{3}x_4+x_{5}+x_6-1)}.
    \label{negQ_cll}
\end{align}
Putting (\ref{negQ_cll}) into (\ref{fq2_cll}), one can get
\begin{align}
f_{Q}= 1+\frac{\alpha\left(-(-1+x_2(x_3-1)+x_5+x_6)\alpha \pm \sqrt{\alpha^{2} (-1+x_2(x_3-1)+x_5+x_6)^2-16\Lambda x_3(-1+x_3x_4+x_5+x_6)}\right)}{8\Lambda x_3},
\end{align}
Utilizing (\ref{fq2_cll}) and (\ref{negQ_cll}) in (\ref{Gamma}), we have
\begin{align}
   \Gamma =-2+\frac{(x_2(x_3-1)-1+x_5+x_6) \pm \sqrt{ (-1+x_2(x_3-1)+x_5+x_6)^2-16c x_3(-1+x_3x_4+x_5+x_6)}}{2(x_{3}x_4+x_{5}+x_6-1)}.
\end{align}
where $c=\frac{\Lambda}{\alpha^2}$. For $\Gamma$ to be real-valued i.e $\left(x_6>1-x_3x_4-x_5 \land c \geq \frac{(-1+x_2(x_3-1)+x_5+x_6)^2}{16x_3(-1+x_3x_4+x_5+x_6)}\right)$. The dynamical system for this example can be expressed as
\begin{align}
x_2' =\; & 2x_2x_3 \notag \\
& -\frac{
    \left(x_2(x_3 - 1) - 1 + x_5 + x_6\right)x_2x_3 
    \pm 
    x_2x_3\sqrt{
        \left(-1 + x_2(x_3 - 1) + x_5 + x_6\right)^2 
        - 16c x_3(-1 + x_3 x_4 + x_5 + x_6)
    }
}{2(x_3 x_4 + x_5 + x_6 - 1)} \notag \\
& - 2x_2 \left( 
    (1 - 3x_4)x_3 
    - \frac{3}{2} \left( x_5(1 + \omega) + x_6 \right) 
\right),
\end{align}
\begin{align}
    x_3'=-3x_3+x_3^2-x_3\left(x_3(1-3x_4)-\frac{3}{2}x_5(1+\omega)-\frac{3}{2}x_6\right),
\end{align}
\begin{align}
    x_4'=1+x_2-3x_4-x_6-x_4 \left(x_3-3x_3x_4 -\frac{3}{2}x_5 (1+\omega)-\frac{3}{2}x_6\right),
\end{align}
\begin{align}
    x_5'=-3x_5 (1+\omega)+x_5 x_3 -2x_5 \left(x_3(1-3x_4)-\frac{3}{2}x_5(1+\omega)-\frac{3}{2}x_6\right),
\end{align}
\begin{align}
    x_6'=2x_3x_6-6x_6-2x_6 \left(x_3-3x_3x_4 -\frac{3}{2}x_5 (1+\omega)-\frac{3}{2}x_6\right).
\end{align}
The physically viable region of this model based on nonnegative conditions of $x_{5}$ and $x_{6}$, and also $f_{Q}>0$ as follows,
\begin{align}\label{vcM3}
\Big( 0 \leq x_6 \leq 1  \land 0\leq x_5 <1-x_6  \land x_3 \leq 0  \land c < 0  \land x_2 < \frac{1 + 8 c x_3 - x_5 - x_6}{ x_3+1}  \land 
  x_4 > \frac{-x_2 - 4 c x_3 + x_2 x_3}{x_2+1}\Big).
\end{align}

\begin{table}[H]
\centering
\begin{tabular}{|p{2cm}|p{5cm}|p{4cm}|}
\hline
\textbf{Fixed Points} & \textbf{$(x_2,x_3,x_4,x_5,x_6)$} & \textbf{Existence}  \\ \hline
$B_1$     & $(0,3,x_4,1-3x_4,1-3x_4) $    & unphysical \\ \hline
$B_2$     & $(x_2,0,\frac{1+x_2}{3},0,0)$ & $ x_2 \in ( \frac{-5 - \sqrt{21}}{2},\; -1)
\cup (\frac{-5 + \sqrt{21}}{2},\; 1)$ \\ \hline
$B_3$    & $(\frac{1+6c}{2},3,x_4,\frac{1-6c-6x_4}{2},\frac{3(1+2c-2x_4)}{2})$      & unphysical \\ \hline
$B_4$    &  $(\frac{1+6c}{2},3,\frac{1+2c}{2},-1-6c,0)$    & unphysical   \\ \hline
\end{tabular}
\caption{Critical points and their existence conditions for non-relativistic matter($\omega=0$) (Model-III). }
\label{table5_cll}
\end{table}


\begin{table}[H]
\centering
\begin{tabular}{|p{2cm}|p{5cm}|p{1.5cm}|p{4cm}|}
\hline
\textbf{Fixed Points} & \textbf{Eigenvalues} & \textbf{Stability}& \textbf{Cosmology}  \\ \hline
$B_2$     & $(-6,-3,-3,-3,0)$       & stable&de Sitter ($H=constant$) \\ \hline
\end{tabular}
\caption{Eigenvalues, stability, and cosmology of the viable fixed points (Model-III).}
\label{table6_cll}
\end{table}
The fixed point analysis for this model is summarized in Table \ref{table5_cll} and Table \ref{table6_cll}. The critical points $B_1$ and $B_3$ form families of anisotropic solutions, but not viable stationary points, as well as the isotropic fixed point $B_4$, as per the viable region of this example mentioned in (\ref{vcM3}). The point $B_2$ representing a family of isotropic solutions serves as a future attractor associated with a de Sitter expansion.

\subsection{Model-IV: $f(Q) = Q e^ {\lambda \frac{Q_{0}}{Q}}$}\label{model4_cll}
The forms of $Q$, $f_Q$, and $\Gamma$ in terms of variables for our study can be obtained as: we have
\begin{align}
    f_Q=e^{\frac{\lambda Q_0}{Q}}\left(1-\frac{\lambda Q_0}{Q}\right), ~~ f_{QQ}=e^{\frac{\lambda Q_0}{Q}}\left(\frac{\lambda^2 Q_0^2}{Q^3}\right).
    \label{fq3_cll}
\end{align}
Explicit substitution into (\ref{29a}) gives
\begin{align}
    Q = \frac{ \lambda Q_{0}(x_2+x_3x_4+x_5+x_6-1)}{x_3x_4+x_5+x_6-1},
    \label{Qe_cll}
\end{align}
substituting  (\ref{Qe_cll}) into (\ref{fq3_cll}), one can get
\begin{align}
f_Q = e^{\frac{x_3x_4+x_5+x_6-1}{x_2+x_3x_4+x_5+x_6-1}} \left( \frac{x_2}{x_2+x_3x_4+x_5+x_6-1} \right),
\end{align}
using (\ref{fq3_cll}) and (\ref{Qe_cll}) into (\ref{Gamma}), one can get
\begin{align}
   \Gamma = \frac{x_2(x_2+x_3x_4+x_5+x_6-1)}{(x_3x_4+x_5+x_6-1)^{2}}.
\end{align}
 We get the dynamical system of this model as
\begin{align}
    x_2'=-\frac{x_2^2x_3(x_2+x_3x_4+x_5+x_6-1)}{(x_3x_4+x_5+x_6-1)^{2}} -2x_2 \left ((1-3x_4)x_3-\frac{3}{2}(x_5(1+\omega)+x_6)\right),
\end{align}
\begin{align}
    x_3'=-3x_3+x_3^2-x_3\left(x_3(1-3x_4)-\frac{3}{2}x_5(1+\omega)-\frac{3}{2}x_6\right),
\end{align}
\begin{align}
    x_4'=1+x_2-3x_4-x_6-x_4 \left(x_3-3x_3x_4 -\frac{3}{2}x_5 (1+\omega)-\frac{3}{2}x_6\right),
\end{align}
\begin{align}
    x_5'=-3x_5 (1+\omega)+x_5 x_3 -2x_5 \left(x_3(1-3x_4)-\frac{3}{2}x_5(1+\omega)-\frac{3}{2}x_6\right),
\end{align}
\begin{align}
    x_6'=2x_3x_6-6x_6-2x_6 \left(x_3-3x_3x_4 -\frac{3}{2}x_5 (1+\omega)-\frac{3}{2}x_6\right).
\end{align}
The physical viable region for this example is given as
\begin{align}\label{vcM4}
\Big( 0 \leq x_6 \leq 1 \land  x_2 \leq 0  \land x_4 > 0  \land  x_3 < \frac{1 - x_2 - x_5 - x_6}{x_4} \land 0 \leq x_5<1-x_6
\Big).
\end{align}

\begin{table}[H]
\centering
\begin{tabular}{|p{2cm}|p{5cm}|p{3cm}|}
\hline
\textbf{Fixed Points} & \textbf{$(x_2,x_3,x_4,x_5,x_6)$} & \textbf{Existence}  \\ \hline
$C_1$     & $(x_2,0,\frac{1+x_2}{3},0,0) $    &$ -1<x_2 \leq 0 $      \\ \hline
$C_2$     & $(0,3,x_4,1-3x_4,1-3x_4)$       & unphysical \\ \hline
$C_3$    & $(x_2,\frac{3}{2},\frac{2(1-x_2)}{3}, 0,2x_2)$      &unphysical  \\ \hline
$C_4$    &  $(x_2,3,\frac{1+x_2}{3},-2x_2,0)$    & unphysical   \\ \hline
\end{tabular}
\caption{Critical points and their existence conditions for non-relativistic matter($\omega=0$) (Model-IV).
 }
\label{table7_cll}
\end{table}

\begin{table}[H]
\centering
\begin{tabular}{|p{2cm}|p{5cm}|p{1.5cm}|p{4cm}|}
\hline
\textbf{Fixed Points} & \textbf{Eigenvalues} & \textbf{Stability}  & \textbf{Cosmology}  \\ \hline
$C_1$     & $(-6,-3,-3,-3,0) $    &stable&  de Sitter ($H=constant$)   \\ \hline

\end{tabular}
\caption{Eigenvalues, stability, and cosmology of the viable fixed points (Model-IV).}
\label{table8_cll}
\end{table}
This model exhibits two families of anisotropic fixed points, namely $C_2$ and $C_3$, also the family of isotropic critical points $C_4$, are not viable, as observed in the previous examples. The viable region of this model is provided in (\ref{vcM4}). The family of isotropic FLRW stationary points $C_1$ represents the de Sitter universe. A detailed summary of these critical points is provided in Table \ref{table7_cll} and Table \ref{table8_cll}, respectively.

\section{Application in the context of early universe cosmology}\label{sec6}

Although spatial anisotropy in the late-time observable universe is tightly constrained by the observations, this does not preclude the possibility of non-negligible anisotropies in the early universe. Any successful early universe cosmological model must contain an isotropization mechanism. Within GR, typically, small spatial anisotropy evolves as $\sim\frac{1}{a^6}$, as can be seen from \eqref{3eq:m}. Small anisotropy therefore, does not really pose a problem for the inflationary models, which is typically modelled with a slow-rolling scalar field. In a pre-bounce ekpyrotic contracting phase $a(t)\sim(-t)^n$ ($t<0,\,0<n<1$), however, any initially existing small classical anisotropy grows as $\sim\frac{1}{a^6}$, faster (as $a\rightarrow 0$) than the energy density of any reasonable matter component, say, with energy density $\rho\sim\frac{1}{a^{3(1+\omega)}}$ with barotropic equation of state (e.o.s.) parameter $\omega=\frac{P}{\rho}$ satisfying $-1<\omega<1$. This is the well-known anisotropy problem in General Relativistic nonsingular bouncing models, which requires inclusion of a super-stiff fluid \cite{Erickson:2003zm, Barrow:2015wfa}. An early universe cosmology based on modified gravity theories must also include an isotropization mechanism. The isotropization property for the de-Sitter solution in $R^2$ gravity has been shown explicitly in \cite{Chakraborty:2018bxh}, and the same in more general quadratic gravity is shown in \cite{Barrow:2006xb}.

This section is aimed at investigating the similar isotropization properties for $f(Q)$ theories. One is typically interested in the behaviour of small homogeneous anisotropic perturbations on an otherwise homogeneous background, which can be treated within the Bianchi-I framework (appendix \ref{app}). The key here is to investigate the stability of the isotropic invariant submanifold. To be non-biased, the isotropization mechanism must be investigated separately in the context of the two different contender paradigms of the early universe, namely inflation and nonsingular bounce.

\subsection{Isotropization during inflation}

In the context of inflation, it suffices to consider a vacuum scenario, since in a modified gravity-based inflationary scenario, there is no additional inflaton field; the additional dynamical degree of freedom coming from the gravity sector is responsible for driving the inflation.
\begin{itemize}
    \item \textbf{Coincidence gauge $\Gamma_1$:} One can simply consider the flow along the vacuum invariant submanifold $x_1=0$ of the dynamical system \eqref{gds1}-\eqref{gds2}:
    \begin{equation}\label{DS1}
    x'_{3} = 6x_{3}(x_3 - 1)\,.
    \end{equation}
    Since $x_3=0$ is a stable solution to the above equation, it turns out that the isotropization dynamics during inflation in the coincident gauge is independent of the underlying theory, since the quantity $\Gamma=\frac{f_Q}{Q f_{QQ}}$ does not play any role in Eq.\eqref{DS1}. An inflating cosmology in this scenario generically isotropizes.
    \item \textbf{Non-coincidence gauge $\Gamma_2$:} In this case, one must consider the dynamics in the vicinity of the invariant submanifold $x_6=0$ within the vacuum invariant submanifold $x_5=0$. The $x_6'$ equation \eqref{x6'} can be conveniently written as
    \begin{equation}
        x_6' = -2x_6(3-\epsilon-x_3),
    \end{equation}
    where $\epsilon=\frac{\dot{H}}{H^2}$ is the standard Hubble slow-roll parameter introduced to characterize the inflationary kinematics. Since $0<\epsilon\ll1$ during inflation, it turns out that the isotropization dynamics depends on whether $x_3<3$ or $x_3>3$. Since $x_3=-\frac{1}{\Gamma}\frac{\dot{Q}}{QH}$, the isotropization dynamics for the non-coincident gauge actually depends on the particular model.
\end{itemize}

\subsection{Isotropization during a pre-bounce ekpyrotic contraction}

In the context of the non-singular bouncing paradigm, the isotropization actually take place during the pre-bounce ekpyrotic contraction phase, which is a slow power law contraction phase. In GR, one typically needs an effective ultra-stiff fluid ($\omega>1$) for this. Whether the additional dynamical degree of freedom presented by the modified gravity theory can do the job without needing an ultrastill fluid is an interesting question. A similar question in the context of $f(R)$ gravity has been addressed in \cite{Arora:2022dti}.

During the pre-bounce contracting phase, one cannot ignore the matter contribution, since it actually grows with time. Therefore, one needs to consider the dynamics in the vicinity of the isotropic invariant submanifold in the presence of matter. Moreover, the appropriate dimensionless time variable on the phase space has to be taken as $N=-\ln a$ for a contracting universe (see e.g.  \cite{Arora:2022dti}); so that all the dynamical equations assume an overall minus sign on the right-hand side. 
\begin{itemize}
    \item \textbf{Coincidence gauge $\Gamma_1$:} The dynamics of the isotropic invariant submanifold during the contraction phase is given by
    \begin{equation}
        x_3' = 6x_3(1 - x_3 - (1+\omega)x_1).
    \end{equation}
    The isotropization depends on whether $x_1$ as well as $\omega$. As long as $(1+\omega)x_1>1$, a pre-bounce ekpyrotically contracting universe isotropizes. The isotropization dynamics depends on the dominant fluid of the universe as well as the underlying theory (since $f_Q$ enters the definition of $x_1$).
    \item \textbf{Non-coincidence gauge $\Gamma_2$:} The dynamics of the isotropic invariant submanifold during the contraction phase is given by the $x_6'$ equation \eqref{x6'} with an inverted sign of the right hand side
    \begin{equation}
        x_6' = -2x_6\left(-\frac{\dot{H}}{H^2} - 3 + x_3\right).
    \end{equation} 
    Typically, an ekpyrotic contraction phase is characterized by a slow power law contraction phase $a(t)\sim(-t)^{\alpha}$ ($0<\alpha\ll1$, $t<0$). Then the above equation becomes
    \begin{equation}
         x_6' = -2x_6\left(\frac{1}{\alpha} - 3 + x_3\right).
    \end{equation}
    Even though $x_3=-\frac{1}{\Gamma}\frac{\dot{Q}}{QH}$ is a model-dependent quantity, if the contracting phase is slow enough ($0<\alpha\ll1$), it isotropizes. The isotropization of a pre-bounce ekpyrotic contraction universe for the non-coincident gauge of $f(Q)$ under consideration is, therefore, although not completely independent, a very likely scenario.
\end{itemize}
We note that although nonsingular bouncing solutions have previously been explored in some works \cite{q3,Koussour:2024wtt,Bajardi:2020fxh}, none of these works have investigated how small anisotropy behaves during a pre-bounce contraction phase.

\section{Comparison with $f(T)$ gravity}\label{comptfTG}

Since the non-metricity theory with connection class I yields identical dynamics as the metric teleparallel gravity, so in this section we briefly compare our findings with some existing studies in $f(T)$ theory. The models \ref{model2} and \ref{model4} were introduced in metric telelparallelism format, $f(T)=T+f_0T^n, ~f(T)=T+f_0(1-e^{-pT^{n}})$ in \cite{fTmodels}. In \cite{comp1}, authors analyzed the Kasner universe \footnote{A special case of the Bianchi I spactime in vacuum with power-law scale factors, i.e., $a_i(t)=t^p_i$:
$$ds^2=-dt^2+t^{2p_1}dx^2+t^{2p_2}dy^2+t^{2p_3}dz^2$$
with two constraints: $p_1+p_2+p_3=1, ~ p_1^2+p_2^2+p_3^2=1$.\\
Although investigations of Kasner solutions in teleparallel gravity provide useful insights into exact anisotropic vacuum configurations, such approaches are inherently limited in their physical scope compared to dynamical analyses in full Bianchi type I spacetimes. The Kasner solution represents a very special, non-generic class of anisotropic evolution characterized by a fixed power-law scale factor behavior. Crucially, it lacks the richness needed to address key cosmological questions such as isotropisation, due to the rigid constraint structure and absence of matter that could drive such behavior. In contrast, Bianchi type I cosmology allows for the inclusion of matter sources, scalar fields, and dynamical degrees of freedom, making it a more general and realistic framework for studying early-universe anisotropies and their suppression during inflation or bounce phases. } in these two particular $f(T)$ gravity models. The authors specifically focused on the cases which correspond to the models we studied in \ref{model2} and \ref{model4}. Their findings revealed an unstable Kasner point for both models, which aligns with our results, that is, the only anisotropic point in both models gives unstable behavior for $\omega<-1$. On the other hand, it was refined in \cite{comp2} that for the power law model, i.e, $f(T)=T+f_0T^N$ except when $T=0$, i.e., except at GR limit, Kasner solution is an unstable asymptotic solution. Moreover, this represents a source in an expanding universe, implying that it was stable in the past as well as an attractor for a contracting universe. When $T=0$, Kasner solution is the general vacuum solution in $f(T)$ gravity, when $T=T_0$ is a non-zero constant, the field equation reduces to GR with a cosmological constant $\Lambda(T_0)$ \cite{adcqg}, giving us no new insight.\\\\


\section{Conclusion}\label{sec7}


In this paper, we have considered the Bianchi-I cosmology in $f(Q)$ theory, conducted the general dynamical system analysis, and discussed its application in the context of late-time and early universe cosmology for connection branch $\Gamma_1$ and $\Gamma_2$. In this study, we have first formulated the general dynamical system, independent of any specific choice of function for $f(Q)$. Thereafter, we have analyzed the dynamical systems of four different $f(Q)$ gravity models: $f(Q) =\alpha (-Q)^{n}$, $f(Q) =Q+\alpha Q^{2}$, $f(Q)= Q+ \alpha \sqrt{-Q}+ \Lambda$ and $f(Q)=Qe^{\lambda \frac{Q_{0}}{Q}}$. For $\Gamma_1$ connection, the first power law model, we get a one-dimensional dynamical system with two critical points, O and P. The critical point P represents matter dominance universe in the dust case and stability behavior varies as given in Table \ref{tablenew2}, which is consistent with the analysis has been studied in \cite{Esposito:2022omp}. The fixed point O depicts a Kasner like solution with stability scenario as mentioned in Table \ref{tablenew2}.  We have noticed that the quadratic model $f(Q) =Q+\alpha Q^{2}$ is independent of the model parameter $\alpha$. The fixed point analysis shows that O, P, and Q are FLRW fixed points, while R and S are anisotropic. We have imposed the strictly positive condition of $f_{Q}$ and non-negative condition of $x_{1}$ and $x_{3}$ ( i.e $f_{Q}>0$, $x_{1} \geq 0$ and $x_{3} \geq 0$). For this model, (\ref{1MC}) governs the physical viability of the stationary points, which are O, P, and S, as shown in Fig. \ref{fig1}. Q and R appear as unphysical points for $\omega=0$. The anisotropic fixed point S describes a Kasner-like solution. The phase portrait in Fig \ref{fig1} shows that O and P lie in the accelerated epoch of the universe. The only saddle point indicating the matter-dominated phase is P, but we have found that for $\omega=0$, it does not provide matter-dominated cosmology, e.g., $a \sim t^{\frac{2}{3}}$. We have concluded that the quadratic model $f(Q)= Q+\alpha Q^{2}$ does not flare well with anisotropy, at least for the late time cosmology and for the connection branch considered.\\\\
The dynamical system of the third model $f(Q) = Q + \alpha \sqrt {-Q} +\Lambda$ depends on the model parameters $\alpha$ and $\Lambda$. For this model, we have four critical points $O_{1}$, $P_{1}$, $R_{1}$, $S_{1}$; among which $S_{1}$ is the only anisotropic fixed point.
For the phase portrait to be real-valued, the range of $c$ has been mentioned in (\ref{condition:c}). The physical viable region for this model is given in (\ref{2MVR}), by setting the conditions $f_{Q}>0$, $x_{1} \geq 0$, and $x_{3} \geq 0$. Complete analyses of fixed points are discussed in Table \ref{table3} and \ref{table4}. For $c=-\frac{1}{4}$, we have four viable fixed points, and for the remaining range of $c$, another three, as shown in Fig \ref{fig2} and Fig \ref{fig3} respectively. The behavior of the fixed points in the real phase portrait stays almost the same for the entire range of $c$ except at $c=-\frac{1}{4}$. The anisotropic fixed point $S_{1}$ gives a Kasner-like solution. For this model, the future attractor $O_1$ serves as the dark energy dominated epoch whereas $P_{1}$ is the only saddle point showing the matter-dominated epoch for $\omega=0$ e.g $a \sim t^{\frac{2}{3}}$, implies the same evolution of the universe as in GR. This model gives asymptotic isotropization in late time cosmology, making it a stable model against small homogeneous anisotropic perturbations in the spacetime geometry. However, this remark is independent on the sign or the value of the model parameter $\alpha$.\\\\
The fourth model $f(Q)=Qe^{\lambda \frac{Q_{0}}{Q}}$ stays independent of the model parameter $\lambda$. Critical point analysis reveals that $O^{'}$, $P^{'}$ and $Q^{'}$ represent the isotropic FLRW points; however, $S^{'}$ is the only anisotropic fixed point. The physical viable region obtained in (\ref{3MVR}), implementing the same conditions on $f_{Q}$, $x_{1}$ and $x_{3}$ as in the previous two models, indicates that all points are physically viable. The $S^{'}$ has a Kasner-like solution similar to those observed in the previous models. The future attractors $O^{'}$ and $Q^{'}$ serve as the dark energy-dominated epoch. The saddle point $P^{'}$ shows the matter-dominant epoch for $\omega= 0$ (dust), by providing the cosmology, i.e. $a \sim t^{\frac{2}{3}}$. Similar to the third model, this model also gives asymptotic isotropization in late time cosmology, making it a stable model against small homogeneous anisotropic perturbations in the spacetime geometry. However, this remark is independent on the sign or the value of the model parameter $Q_0$ or $\lambda$.\\\\
In the framework of early universe cosmology, our analysis reveals that the dynamical behavior of Bianchi-I spacetimes within the connection branch $\Gamma_1$ exhibits a model independent tendency toward isotropization in the vacuum case. Specifically, we have concluded that inflationary evolution invariably drives the universe toward isotropy, regardless of the underlying $f(Q)$ theory. In contrast, pre-bounce ekpyrotic contracting universe isotropization depends on the dynamics of the dominant fluid of the universe as well as the underlying theory.\\\\
For $\Gamma_2$ connection, the power-law model reduces to a four-dimensional dynamical system upon imposing an additional constraint, as specified in (\ref{anotherconstr_cll}). This particular setup admits five isotropic FLRW critical points, as well as a family of anisotropic critical points. However, based on the viability conditions, we find that five fixed points, comprising four isotropic and one family of anisotropic points, are physically unviable. The only viable fixed point corresponds to a future de Sitter attractor solution, as outlined in Table \ref{table1_cll} and \ref{table2_cll}.
The quadratic model $f(Q) =Q+\alpha Q^{2}$ yields a family of isotropic solutions that corresponds to a future de Sitter universe. However, the anisotropic families of critical points do not satisfy the physical viability conditions, as shown in Table \ref{table3_cll} and \ref{table4_cll}.\\\\
The dynamical system of the third model $f(Q) = Q + \alpha \sqrt {-Q} +\Lambda$ depends on the model parameters $\alpha$ and $\Lambda$. For this model, we have four critical points. Complete analyses of fixed points
are discussed in Table \ref{table5_cll} and \ref{table6_cll}. The critical points $B_1$ and $B_3$ form families of anisotropic solutions but lie outside the physically viable region. The point $B_2$ representing a family of isotropic solutions serves as a future attractor de Sitter solution, whereas $B_4$ is also unphysical. Our final exponential model also gives four critical points. A detailed summary of these critical points is provided in Table \ref{table7_cll} and \ref{table8_cll}, respectively.  Among them, two are families of anisotropic fixed points, labeled $C_2$ and $C_3$, which are unphysical, as observed in the previous models. However, the family of isotropic FLRW points $C_1$ corresponds to the de Sitter solution. The fixed point $C_4$ is also not viable.\\\\
In the context of early universe cosmology within the connection branch $\Gamma_2$, our analysis demonstrates that the dynamical evolution of Bianchi-I spacetimes in vacuum exhibits a model-dependent tendency toward isotropization. Specifically, we find that both inflationary expansion and pre-bounce ekpyrotic contraction phases generically drive the universe toward isotropy, subject to the specific form of the underlying $f(Q)$ gravity model.\\\\
Hence, in general, we can make the following statements based on our analysis: Kasner solutions, which appear in all the models under consideration in both connection branches, marginally violate the $f_Q>0$ condition. A physically viable stable de-Sitter attractor is a generic feature, except for the very special case of the monomial model $f(Q)\propto(-Q)^n$ within the coincident gauge connection. The isotropization of a homogeneously perturbed inflating FLRW universe is a generic model-independent feature in the coincident gauge, and the isotropization of a homogeneously perturbed pre-bounce ekpyrotically contracting FLRW universe is, although not completely generic, a likely scenario.

\appendix

\section{Bianchi-I cosmology as a homogeneous perturbation over spatially flat FLRW}\label{app}

In the main text, we have considered Bianchi-I cosmology in the presence of an isotropic fluid. In other words, we have considered anisotropy in the geometry sector, but no anisotropy in the fluid sector. Although many works in the literature consider a similar situation, one may wonder, and rightfully so, how justified this is. Such a situation can be strictly justified if the anisotropy is initially taken as perturbatively small, although it can definitely grow along the course of the evolution. Below we show that the Bianchi-I spacetime can be considered as a homogeneously perturbed FLRW spacetime in the synchronous gauge. The discussion is borrowed from \cite{Arora:2022dti}.

Consider the spatially flat FLRW cosmology, which assumes homogeneity and isotropy
\begin{align}
    ds^2=-dt^2+a^2(t)\delta_{ij}dx^i dx^j.
\end{align}
Small-scale deviations from homogeneity and isotropy in the cosmological background are treated as cosmological perturbations. The perturbed spatially flat FLRW metric is expressed as
\begin{align}
    ds^2 = -(1 + 2\varphi(\bar{x},t)) dt^2 - 2C_i(\bar{x},t) dtdx^i  
    + a^2(t) \left(1 - 2\Psi(\bar{x},t) \delta_{ij} + 2h_{ij}(\bar{x},t) \right) dx^i dx^j,
\end{align}
where $\varphi, C_i, \Psi$ and $h_{ij}$ are perturbation quantities that depend on spacetime. $\Psi$ represents trace and $h_{ij}$ denotes the traceless part of $\delta g_{ij}$. In the synchronous gauge, the above metric reduces to
\begin{align}
    ds^2=-dt^2 +a^2(t)(1-2\Psi (\bar{x},t)\delta_{ij}+2h_{ij}(\bar{x},t))dx^i dx^j.
\end{align}
In this study, we have neglected any inhomogeneity and just focused on small anisotropy. Therefore we can write
\begin{align}
    ds^2=-dt^2 +a^2(t)(1-2\Psi (t)\delta_{ij}+2h_{ij}(t))dx^i dx^j.
\end{align}
Let $\beta_{1}(t)$, $\beta_{2}(t)$ and $\beta_{3}(t)$ are the eigenvalues of order three matrix $h_{ij}(t)$. As we already mentioned that $h_{ij}(t)$ is traceless, this implies $\beta_{1}+\beta_{2}+\beta_{3}=0$. By transforming the basis $x^i$ to eigenvector basis of $h_{ij}(t)$, we can express the spatial part of perturbed metric as
\begin{align}
    g_{ij}(t)=a^2(t)\left (1-2\Psi (t)\delta_{ij}+diag(2\beta_{1}(t),2\beta_{2}(t),2\beta_{3}(t))\right ),
\end{align}
Alternative, since $\beta_{i}$s and $\Psi$ are perturbation quantities, we have
\begin{align}
    g_{ij}(t)=a^2(t)diag \left (e^{-2\Psi(t)+2\beta_{1}(t)},e^{-2\Psi(t)+2\beta_{2}(t)},e^{-2\Psi(t)+2\beta_{3}(t)}\right ).
\end{align}
Now, by absorbing trace part $e^{-2\Psi(t)}$ into $a(t)$, we can write the parametrization form of spatially flat homogeneous and anisotropic Bianchi-I metric as
\begin{align}
    ds^2=-dt^2+a^2(t) \left (e^{2\beta_{1}(t)}dx_{1}^2+e^{2\beta_{2}(t)}dx_{2}^2+e^{2\beta_{3}(t)}dx_{3}^2\right ).
\end{align}\\\\

\section{Derivation of anisotropy evolution}\label{dotsigma}
From the Friedmann type equations of $f(Q)$ gravity in Bianchi I spacetime with connection class $\Gamma_{II}$, we originally have \footnote{$\gamma=0$ reduces this to the formulation of connection class $\Gamma_I$}
\begin{align}
    \kappa p_i=-\frac{f}{2}+(H_i-3H+\frac{3 \gamma}{2})\dot{f_Q}+(\dot{H}_i-3\dot{H}+3HH_i-9H^2)f_Q.
\end{align}
So
\begin{align}
    \kappa \sum_i p_i=-\frac{3f}{2}+(3H-9H+\frac{9 \gamma}{2})\dot{f_Q}+(3\dot{H}-9\dot{H}+9H^2-27H^2)f_Q.
\end{align}
Since we consider isotropic fluid, we can express $p_i=\frac{1}{3} \sum_l p_l$, therefore from the above equations we obtain
\begin{align}\label{betaeq}
    0=\ddot{\beta_i}f_Q+\dot{\beta_i}(\dot{f_Q}+3Hf_Q).
\end{align}
Since
\begin{align*}
    \sigma^2=\sum_l\dot{\beta^2_l},
\end{align*}
we can write
\begin{align}\label{sig}
    \frac{\dot{\sigma}}{\sigma}=\frac{\sum_l\ddot{\beta_l}\dot{\beta_l}}{\sum_l\dot{\beta^2_l}}.
\end{align}
Using (\ref{sig}) and after some simplification, (\ref{betaeq}) can be expressed as
\begin{align*}
    0=\frac{\dot{\sigma}}{\sigma}+\frac{\dot{f_Q}}{f_Q}+3H.
\end{align*}

\end{document}